%% file: main.tex
\documentclass[conference]{IEEEtran}
\IEEEoverridecommandlockouts
\usepackage{cite}
\usepackage{amsmath,amssymb,amsfonts}
\usepackage{algorithmic}
\usepackage{graphicx}
\usepackage{pbox}
\usepackage{textcomp}
\usepackage{xcolor}
\usepackage{url}
\usepackage{multirow}
\usepackage{booktabs}
\usepackage[normalem]{ulem}
\usepackage{listings}
\makeatletter
\newcommand{\linebreakand}{%
  \end{@IEEEauthorhalign}
  \hfill\mbox{}\par
  \mbox{}\hfill\begin{@IEEEauthorhalign}
}
\makeatother
\definecolor{jsonkey}{rgb}{0.75, 0.2, 0.2}
\definecolor{jsonstring}{rgb}{0.25, 0.5, 0.35}

\lstdefinelanguage{json}{
  basicstyle=\ttfamily\footnotesize,
  showstringspaces=false,
  breaklines=true,
  frame=single,
  backgroundcolor=\color{gray!10},
  stringstyle=\color{jsonstring},
  commentstyle=\color{gray},
  morecomment=[l]{//},
  morecomment=[s]{/*}{*/},
  morekeywords={true,false,null},
  keywordstyle=\color{jsonkey}\bfseries
}

\lstdefinestyle{jsonstyle}{
  language=json,
  numbers=left,
  numberstyle=\tiny\color{gray},
  stepnumber=1,
  columns=fullflexible
}

\usepackage{soul}
\colorlet{soulblue}{blue!20}

\def\BibTeX{{\rm B\kern-.05em{\sc i\kern-.025em b}\kern-.08em
    T\kern-.1667em\lower.7ex\hbox{E}\kern-.125emX}}
\begin{document}

\title{RA-QA: A Benchmarking System for \\Respiratory Audio Question Answering Under\\Real-World Heterogeneity
}

\author{\IEEEauthorblockN{1\textsuperscript{st} Gaia A. Bertolino}
\IEEEauthorblockA{\textit{University of Cambridge}\\
Cambridge, United Kingdom\\
gab62@cam.ac.uk}
\and
\IEEEauthorblockN{2\textsuperscript{nd} Yuwei Zhang}
\IEEEauthorblockA{\textit{University of Cambridge}\\
Cambridge, United Kingdom\\
yz798@cam.ac.uk}
\and
\IEEEauthorblockN{3\textsuperscript{rd} Tong Xia}
\IEEEauthorblockA{\textit{Tsinghua University}\\
Beijing, China\\
tongxia@mail.tsinghua.edu.cn}
\linebreakand 
\IEEEauthorblockN{4\textsuperscript{th} Domenico Talia}
\IEEEauthorblockA{\textit{University of Calabria}\\
Rende, Italy\\
talia@dimes.unical.it}
\and
\IEEEauthorblockN{5\textsuperscript{th} Cecilia Mascolo}
\IEEEauthorblockA{\textit{University of Cambridge}\\
Cambridge, United Kingdom\\
cm542@cam.ac.uk}
}
\maketitle
% \gb{REVIEWS\\
% (I) The purpose is not clear whether it is converting multiple audio corpora to a multimodel RA-QA or development of a screening models using the corpus. \\
% (II) It should have been further experimented on the development of the classification models using state of the art pre-trained models. \\
% (III) In the audio-only condition, all that is mentioned is that an SVM is used. What features? What SVM settings/kernels? Why such a simplistic model?\\
% (IV) If using the term benchmark, then add (i) more models, distinguishing those which are in-domain and which are not (ii) more analysis on the results using the SOTA as a reference (for e.g. could put something on the different modalities, the effect of language, so how different question types are handled)
% }
\begin{abstract}
As conversational multimodal AI tools are increasingly adopted to process patient data for health assessment, robust benchmarks are needed to measure progress and expose failure modes under realistic conditions. Despite the importance of respiratory audio for mobile health screening, respiratory audio question answering remains underexplored, with existing studies evaluated narrowly and lacking real-world heterogeneity across modalities, devices, and question types. We hence introduce the \textbf{Respiratory-Audio Question-Answering (RA-QA) benchmark}, including a standardized data generation pipeline, a comprehensive multimodal QA collection, and a unified evaluation protocol. RA-QA harmonizes public RA datasets into a collection of 9 million format-diverse QA pairs covering diagnostic and contextual attributes. We benchmark general audio-language models as well as domain-specific architectures, establishing reproducible reference points and showing how current approaches fail under heterogeneity.  \\
\end{abstract}

\begin{IEEEkeywords}
Respiratory health; Audio Question Answering; Multimodal Benchmark; Audio-Language Models; Healthcare.
\end{IEEEkeywords}

\input{sections/1_introduction}
\input{sections/2_related_work}
\input{sections/3_generation_pipeline}
\input{sections/4_benchmarking}
\input{sections/5_results_old}

\input{sections/6_conclusions}

\input{appendix/datasets}

\input{appendix/construction}

\input{appendix/clinician}
\input{appendix/metrics}
\input{appendix/baselines}

\input{appendix/exp_setting}

\input{appendix/results}

\end{document}

%% file: sections/1_introduction.tex
% \yz{I get your point of adding the usecase in intro. But this does not flow naturally? Why do you start by giving a definition to Conversational AI? } \gb{(i) moved the usecases later (ii) grounding better what is conversational AI may help the reader to understand what is the topic / scope of the paper}
% \yz{And why are the usecases not related to respiratory audio? respiratory and audio as your main focus are introduced too late?} \gb{addressed}
% \yz{Why do you start by \textbf{giving a definition} to Conversational AI? Do you mean to say the recent development of conversational AI enables xxx?}
% \yz{I personally don't like the intro starting sentences and would prefer the abstract flow. But your choice.}
\section{Introduction} \label{sec:intro}
Conversational AI in healthcare refers to interactive systems that can interpret user questions, condition their responses on health-relevant information, and provide context-sensitive answers in clinical or patient-facing workflows. In this setting, the goal is to generate fluent textual answers for reliable information access, monitoring, and decision support, while adhering to clinical safety constraints.  
As conversational multimodal models are increasingly used in clinical-facing settings, rigorous and realistic \textit{benchmarks} have become essential to quantify both capabilities and failure modes under conditions that reflect real workflows (e.g., multi-turn, diverse users, and safety-critical criteria)~\cite{b1}. Accordingly, evaluations should stress-test models across \textit{diverse interaction styles} (e.g., question formats) and \textit{real-world conditions} (e.g., heterogeneous contexts and recording devices), rather than assuming a single, static query and a homogeneous input setting~\cite{b2}.

% \yz{what motivates? conversation AI usecase motivate benchmark? Shouldn't it be the safety-crucial nature of healthcare that requires rigorous benchmarking?}\gb{addressed}

% \textcolor{red}{This motivates the need for rigorous and realistic \textit{benchmarks} to quantify both capabilities and failure modes under conditions that reflect real workflows interactions (e.g., diverse users and safety-critical criteria)~\cite{b1}. Accordingly, evaluations should stress-test models across \textit{diverse interaction styles}, such as question formats, and \textit{real-world conditions}, such as heterogeneous contexts and recording devices, rather than assuming a single, static query and a homogeneous input setting~\cite{b2}. .}
\input{figures/chatbot}
Respiratory health is a particularly compelling setting for benchmark-driven evaluation of conversational AI: pulmonary and airway diseases remain a leading cause of global morbidity and mortality, and practical screening and monitoring tools must operate under noisy, heterogeneous real-world conditions, including telemedicine and low-resource deployments~\cite{b3}.

In clinical practice, auscultation is central to respiratory assessment: acoustic signatures such as wheezes, crackles, and abnormal timing patterns provide key diagnostic cues for conditions including asthma, pneumonia, and chronic obstructive pulmonary disease (COPD). Recent advances in machine learning have achieved strong performance for respiratory sound classification from recordings \cite{b4,b5}, and recent efforts have moved toward reusable respiratory foundations and standardized downstream evaluation \cite{b6}, and multimodal audio-text models for respiratory outcome prediction \cite{b7}. However, most prior work treats respiratory audio as a \textit{single-output prediction} problem, producing a predefined label or score per recording (e.g., diagnosis or symptom presence), rather than supporting question-conditioned, conversational answering (Fig.~\ref{fig:chatbot}). 

% \yz{I would maybe instead add the usecases here?}\gb{done} 
In realistic patient-facing and clinical workflows, respiratory assessment is inherently \textit{question-driven}: users and clinicians may ask diverse, context-dependent questions about the same recording, such as verifying a symptom, probing diagnostic consistency, estimating severity, or querying acquisition conditions. Questions also come in multiple answer formats (e.g., binary, multiple-choice, numeric, or free-form). Supporting these interactions requires models that jointly interpret respiratory audio and a natural-language query and produce clinically grounded responses, rather than a single static label (Fig.~\ref{fig:questions}). Potential use cases include: (i) remote monitoring of patients with chronic respiratory conditions, particularly those with limited access to care; (ii) post-discharge follow-up after acute respiratory illness; and (iii) screening or triage support during periods of high respiratory disease burden.

\input{figures/questions} 

While question answering (QA) has been extensively benchmarked for other clinical modalities (EHR, imaging, biosignals), including large-scale resources for text QA \cite{b8,b9,b10} and multimodal QA for imaging and signals \cite{b11,b12,b13}, respiratory audio-grounded QA remains comparatively under-benchmarked. In parallel, general audio-language models \cite{b14,b15,b16} enable broad audio understanding and open generation from paired audio-text inputs, but they are not designed or systematically evaluated for subtle auscultation cues, disease-specific semantics, and distribution shifts typical of real respiratory recordings. Even though recent respiratory QA efforts have advanced, they are limited in scope or availability: CaReAQA/CaReSound \cite{b17} is a valuable step toward medical audio QA, but does not comprehensively cover the variability encountered in respiratory assessment across auscultatory sounds and cough, breathing, and speech modalities, heterogeneous recording conditions and devices, diverse diseases and attributes, and multiple QA formats under a unified evaluation protocol.

To address this gap, we introduce \textbf{Respiratory-Audio Question-Answering (RA-QA)}, a publicly available benchmarking system for respiratory audio QA. RA-QA (i) provides a standardized generation pipeline for QA-pair generation from existing datasets, (ii) harmonizes multiple public respiratory audio corpora into a large-scale collection of 9 million question-answer pairs spanning diagnostic and contextual attributes across modalities and question formats (Figure~\ref{fig:questions}), (iii) benchmarks both traditional audio ML baselines and general audio-language generation models under a common protocol, enabling systematic and reproducible comparison, and (iv) shows that general audio-language models and generic audio benchmarks do not reliably transfer to subtle respiratory cues, and that semantic similarity can remain high despite low task-level correctness, motivating evaluation that jointly reports semantic fidelity and task-level performance. The full generation code and the released QA pairs are available at \url{https://anonymous.4open.science/r/RA-QA-6CB3/}.

%% file: figures/chatbot.tex
\begin{figure}[t]
  \centering
  \includegraphics[width=0.9\linewidth]{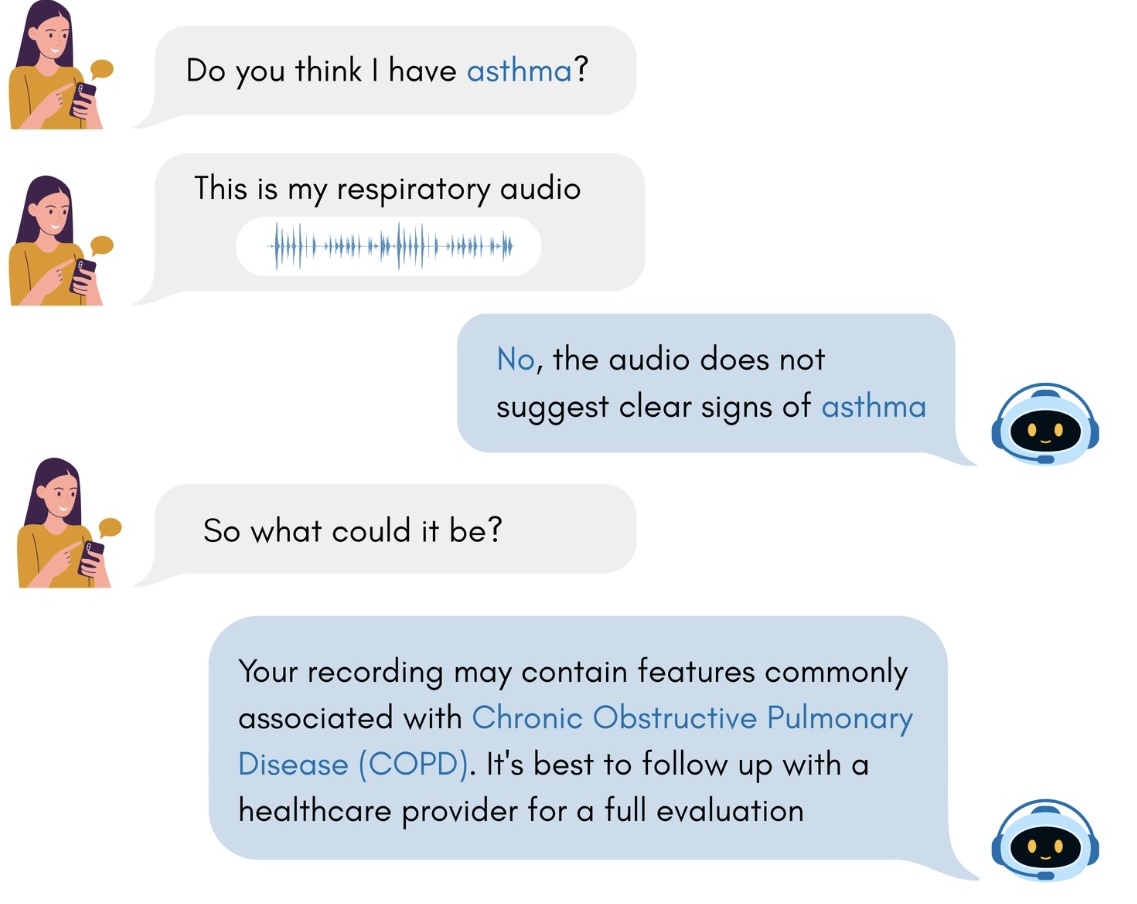}
  \caption{Example scenario of multimodal patient-chatbot interaction. }
  % \yz{did you have a shorter version of this figure? If I remember correctly?}
  % Illustration of a conversational interface where a doctor or patient interacts with an AI assistant by submitting respiratory audio recordings and receiving natural language responses. 
  % The chatbot provides preliminary assessments based on the audio input, supporting clinical decision-making or self-screening.

  \label{fig:chatbot}
\end{figure}

%% file: figures/questions.tex
\begin{figure}[t]
  \centering
  \includegraphics[width=0.9\linewidth]{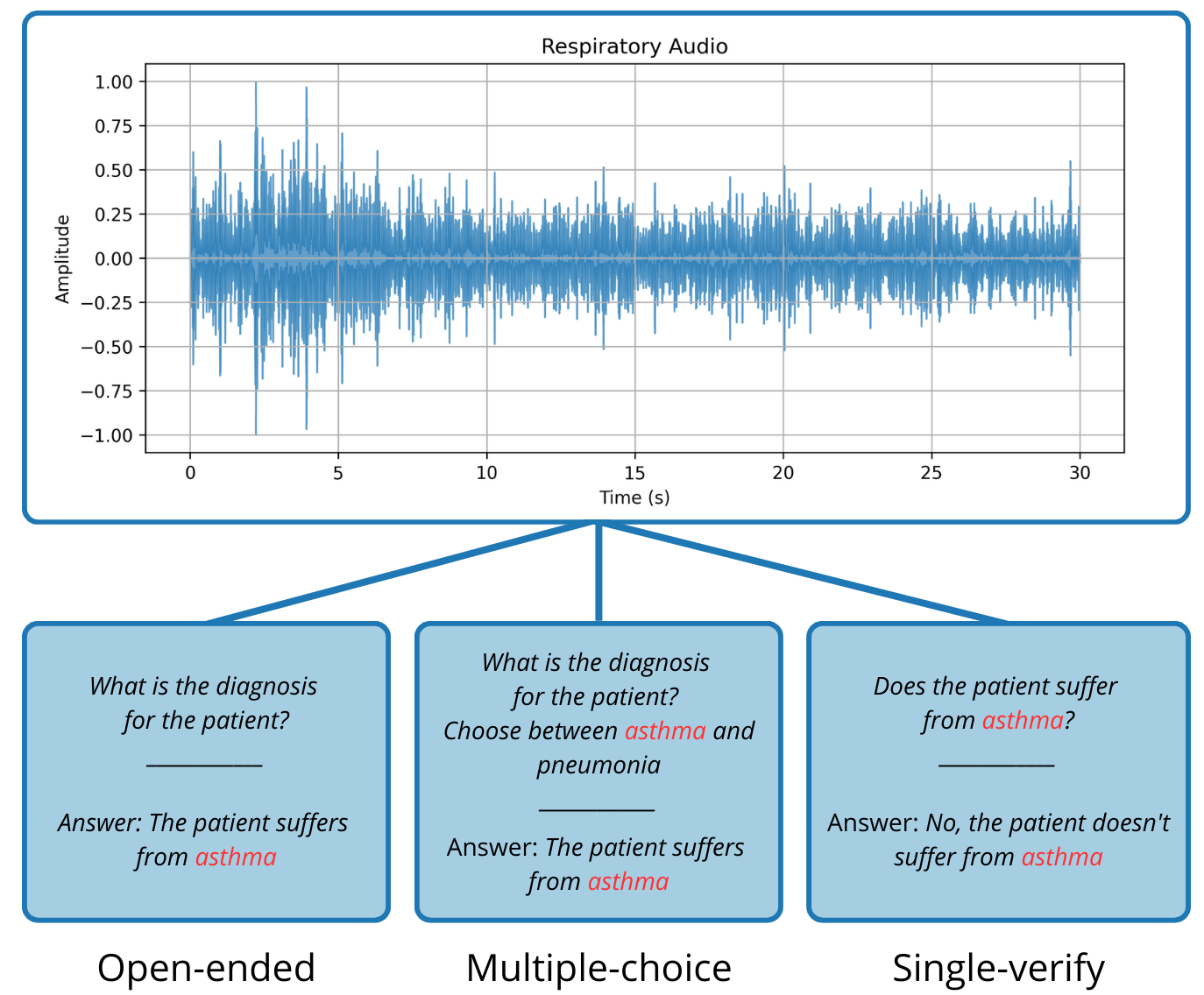}
  \caption{RA-QA diverse question formats.}
  \label{fig:questions}
\end{figure}

%% file: sections/2_related_work.tex
\section{Related work}
\label{sec:related_work}

\textbf{Respiratory audio ML for healthcare.}
A substantial body of work studies respiratory sounds (cough, breath, auscultation) with classical and deep learning pipelines for detection and classification of respiratory conditions and symptoms \cite{b4,b5}. Recent efforts have also moved toward reusable respiratory foundations and standardized downstream evaluation, e.g., OPERA curates large-scale respiratory audio and benchmarks pretrained acoustic backbones across multiple respiratory tasks \cite{b6}. Complementary multimodal approaches such as RespLLM unify audio and text representations for respiratory outcome prediction, largely in a classification-style setup \cite{b7}. However, these lines of work typically optimize for \emph{single, fixed targets} (labels/scores) and do not directly support the interactive, question-driven queries encountered in patient-facing and clinical workflows.

\textbf{Medical QA beyond audio.}
Medical QA has been widely developed in text-centric settings, including QA over clinical notes (emrQA) \cite{b8}, biomedical literature (PubMedQA) \cite{b13}, and consumer medical questions (MedRedQA) \cite{b9}; structured EHR QA has also been explored via text-to-SQL benchmarks such as MIMICSQL and EHRSQL \cite{b31,b32}. In parallel, multimodal medical QA has expanded rapidly for imaging and physiological signals, with benchmarks spanning radiology VQA and broad multi-modality collections (e.g., VQA-Med, OmniMedVQA) as well as signal-grounded QA such as ECG-QA \cite{b11,b12,b10}. These resources demonstrate the value of QA as a unifying interface for heterogeneous clinical data, but respiratory \emph{audio}-grounded QA remains comparatively under-resourced.

\textbf{Audio-language QA and respiratory QA.}
General audio-language models (e.g., Pengi, SpeechVerse, Qwen-Audio) and related audio-instruction frameworks enable broad audio understanding and open-ended generation \cite{b14,b15,b16}. Yet they are not designed or benchmarked for respiratory auscultation cues, disease-specific semantics, and shift conditions typical of real respiratory recordings. The closest direction is emerging \emph{medical-audio} QA, including CaReAQA/CaReSound \cite{b17}, which provides an important first step but remains limited in coverage of respiratory modality diversity (cough/breath/speech), disease/attribute breadth, and multi-dataset heterogeneity under a unified, shift-aware evaluation. In contrast, our RA-QA system is constructed specifically to capture this variability and to enable systematic benchmarking under a common respiratory QA protocol.

%% file: sections/3_generation_pipeline.tex
\section{RA-QA Data Curation} \label{sec:data_generation_pipeline}
The RA-QA pipeline converts heterogeneous respiratory audio datasets into a unified, QA-ready benchmark by harmonizing clinical attributes, standardizing metadata, and automatically generating natural language question-answer pairs. It explicitly operationalizes a dataset-to-QA transformation process, enabling reproducible conversion of diverse respiratory datasets into a unified multimodal QA format. This design ensures consistency, scalability, and interoperability across diverse data sources and clinical contexts. 
Further details are provided in the Appendix~\ref{apd:construction}.

\subsection{RA-QA collection design} \label{sec:design}
The RA-QA benchmark comprises over 9 million question-answer pairs generated from 11 distinct datasets \cite{b18,b19,b20,b21,b22,b23,b24,b25,b26,b27,b28}, encompassing multiple respiratory conditions (e.g., asthma, COPD, COVID-19) and covering a wide range of respiratory audio modalities, including cough, breathing, speech, and auscultation recordings. Table~\ref{tab:overview_collection} summarizes dataset-level statistics, including the number of recordings, QA pairs, and average question/answer lengths (see Figure~\ref{fig:audio_waves}). 
\input{tables/overview_collection}

\input{figures/audio_waveforms}
Several datasets include multiple recordings per participant, collected under different conditions or time points, introducing longitudinal variability that supports patient-specific modeling. Moreover, some datasets provide segment-level labels within the same recording, enabling finer-grained analysis of respiratory events and localized acoustic phenomena. Since some labels are clinically validated and others are self-reported, the resulting diversity in recording conditions and metadata reflects realistic deployment scenarios and supports the development of robust, generalizable respiratory audio QA systems. (See Figure~\ref{fig:distribution}). 
\input{figures/distribution}

We categorize RA-QA targets into four macro-attribute categories that capture complementary clinical signals: (i) acoustic features, including segment-level annotations that support multiple clinically grounded questions per recording and reflect temporal or spatial variability in respiratory sounds; (ii) consultation context, describing symptoms and testing status at the time of recording; (iii) demographics and health profile, providing background clinical context; and (iv) recording context, encoding environmental and procedural factors that influence audio characteristics. Together, these categories encourage models to reason not only over clinically relevant context but also over respiratory acoustics. Meanwhile, RA-QA spans two task families: \textit{discriminative} tasks, where the answer is a categorical label (e.g., diagnosis/symptom presence), and \textit{regression} tasks, where the answer is a continuous numeric value (e.g., physiological measurements).

RA-QA system QA pairs are developed according to three formats: (i) open-ended (OE) questions, which expect free-text responses without predefined options, simulating natural dialogue, (ii) multiple-choice (MC) questions, which follow the same structure as open-ended ones but include a set of suggested answer options within the prompt, following a prompt-engineering approach, and (iii) single-verify (SV) questions, limited to binary responses. The three formats have been developed to increase the contextual and linguistic diversity of the dataset, enabling models, particularly large language models, to better interpret semantically equivalent prompts that may vary in structure or phrasing. This diversity helps simulate real-world scenarios where users (clinicians or patients) might pose questions in different formats. 

\subsection{RA-QA generation pipeline}
A robust pipeline was designed and implemented to uniformly process the diverse datasets (see Figure~\ref{fig:pipeline}) to ensure a harmonized and standardized QA collection, thus achieving consistency across the benchmark. The processing workflow was divided into several stages: 

\noindent \textbf{(i) Metadata Standardization and Label Mapping.}
The first stage of dataset processing focused on standardizing the metadata across the 11 source datasets to ensure consistency in structure and interpretation. As part of the QA formatting process, original categorical labels were systematically mapped to descriptive text strings. While some interpretative decisions were required, they were guided by established clinical terminology and dataset documentation.
\input{figures/pipeline}

\noindent \textbf{(ii) Template Structuring.}
From each standardized metadata file, JSON-based question templates were generated for all relevant attributes. According to the design choice reported in Section~\ref{sec:design}, we defined three question-format templates: open-ended, multiple-choice, and single-verify questions. Templates are instantiated with natural-language question prompts and paired with human-readable answers, such that each instance forms an actual QA exchange rather than a raw label prediction. For regression-valued attributes, we generate open-ended templates only, as enumerating numeric options would have been impractical, whereas for discriminative attributes we instantiate all applicable question types.

% Question types were selected based on attribute structure: binary attributes used single-verify questions, while categorical attributes supported all three formulations, including open-ended and multiple-choice formulations. 
% A harmonization procedure aligned semantically equivalent attributes across datasets by mapping labels to unified, human-readable text strings, ensuring consistent and reproducible question generation).\\

\noindent \textbf{(iii) QA Pair Generation.} Once the templates and standardized metadata were defined, QA pairs were programmatically generated for each individual patient in the datasets. This resulted in a collection of personalized question-answer pairs, where each question targets specific metadata fields (e.g., symptoms, diagnosis, age) and each answer is drawn from the corresponding patient data. These QA pairs are then associated with the patients' respiratory audio recordings, creating a multimodal instance that links textual input, audio context, and textual output. This patient-level generation strategy ensures that the resulting dataset is coherent, scalable, and suitable for training and evaluating multimodal QA systems in respiratory health contexts. Assigned splits followed the original dataset protocols when available; otherwise, stratified splits were applied based on other available attributes. Data were partitioned into 70/15/15 train/validation/test splits.

\subsection{Clinical review of question templates}
To support the clinical validity of RA-QA, the question templates and their candidate answer sets were developed and refined with input from a clinician (see Appendix~\ref{app:clinician-review}).

%% file: tables/overview_collection.tex
\begin{table*}[htbp]
\caption{Overview statistics of the generated respiratory-audio RA-QA collection.}
\vspace{-5pt}
\label{tab:overview_collection}
\centering
\setlength{\tabcolsep}{3pt}

\resizebox{\textwidth}{!}{%
\begin{tabular}{lrrrrrrrrrrrr}
\toprule
 & \textbf{KAUH} & \textbf{Coswara} & \textbf{CoughVID} & \textbf{COVID-19 Sounds} & \textbf{HF Lung} & \textbf{ICBHI} & \textbf{MM Lung} & \textbf{NoseMic} & \textbf{Resp@TR} & \textbf{SSBPR} & \textbf{UK COVID-19} & \textbf{All} \\
\midrule
Number of Audio Samples & 336 & 24860 & 1855 & 2046 & 9765 & 6895 & 80 & 1297 & 504 & 7569 & 213492 & 268699 \\
Mean duration (s) & 17.40 & 9.85 & 8.37 & 14.33 & 15.00 & 2.70 & 7.77 & 30.00 & 21.75 & 2.36 & 6.29 & 6.86 \\ 
\midrule
Mean Question Length & 11.02 & 7.71 & 7.83 & 8.62 & 11.06 & 9.72 & 8.34 & 7.50 & 10.47 & 10.00 & 8.79 & 8.78 \\ 
Number of Unique Questions & 30 & 38 & 37 & 47 & 9 & 25 & 10 & 6 & 12 & 12 & 48 & 207 \\
\midrule
Mean Answer Length & 6.88 & 6.71 & 5.99 & 8.24 & 8.49 & 5.58 & 6.12 & 6.17 & 7.96 & 7.97 & 4.90 & 5.09 \\
Number of Unique Answers & 136 & 146 & 164 & 637 & 20 & 243 & 153 & 620 & 29 & 32 & 232129 & 234064 \\ 
\midrule
\textbf{QA Pairs} & 9033 & 430473 & 76450 & 59187 & 55255 & 171849 & 680 & 7782 & 5544 & 90828 & 8089056 & 8996137 \\
\bottomrule
\end{tabular}%
}

\end{table*}

%% file: figures/audio_waveforms.tex
\begin{figure}[htbp]
  \centering
  \includegraphics[width=1\linewidth]{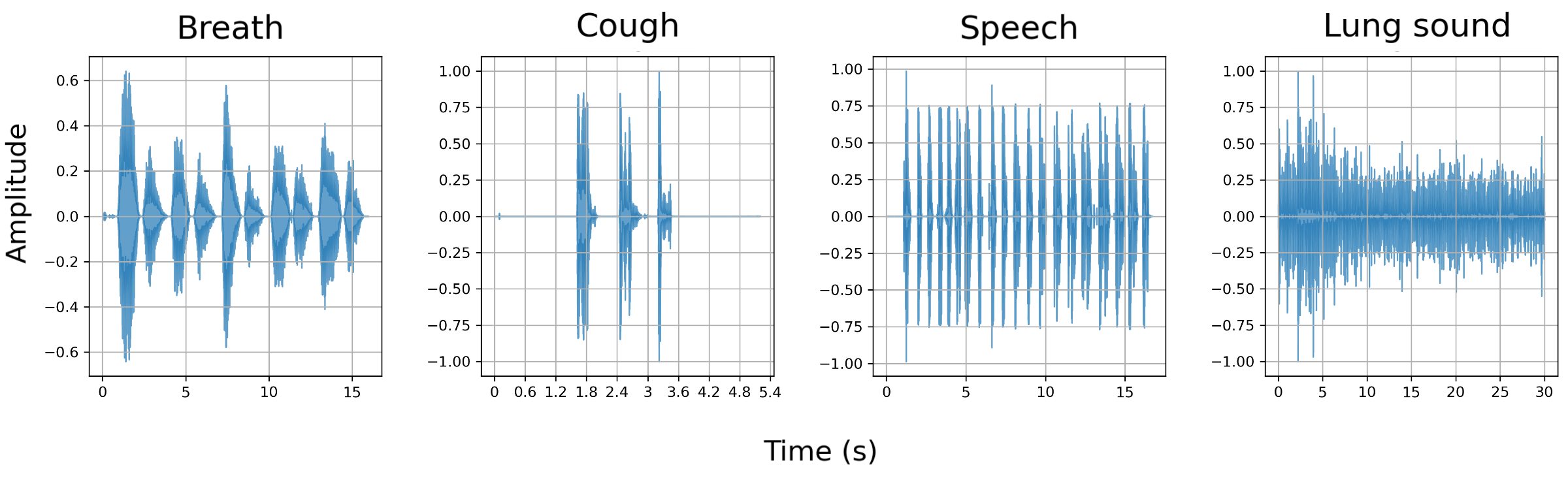}
  \caption{Examples of sound waves in the datasets.}
  \label{fig:audio_waves}
\end{figure}

%% file: figures/distribution.tex
% \begin{figure*}[htbp]
%   \centering
%   \includegraphics[width=\linewidth]{figures/distribution.png}
%   \caption{ \textcolor{red}{Global acoustic density maps computed over the unified latent spaces of OPERA-CT, OPERA-GT, and OPERA-CE via deep-embedding PCA.}}
%   \label{fig:distribution}
% \end{figure*}

\begin{figure}[htbp]
  \centering
  \includegraphics[width=\linewidth]{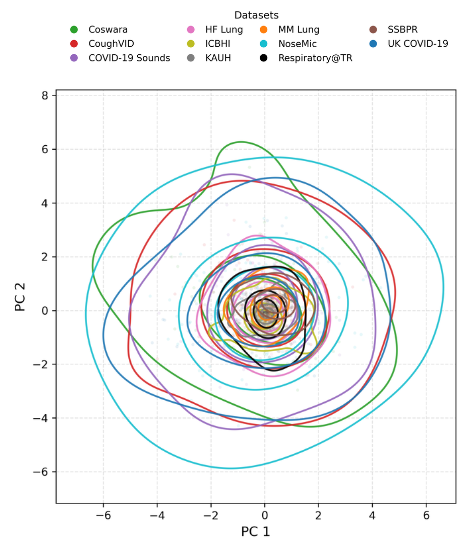}
 \caption{Acoustic density map of RA-QA audio embeddings extracted with the CT encoder from the OPERA framework projected into a 2D PCA space. 
 % \yz{Can you remove the box of the legend and put it above the figure? try to align the width?}
 }
  \label{fig:distribution}
\end{figure}

%% file: figures/pipeline.tex
\begin{figure}[htbp]
  \centering
  \includegraphics[width=\linewidth]{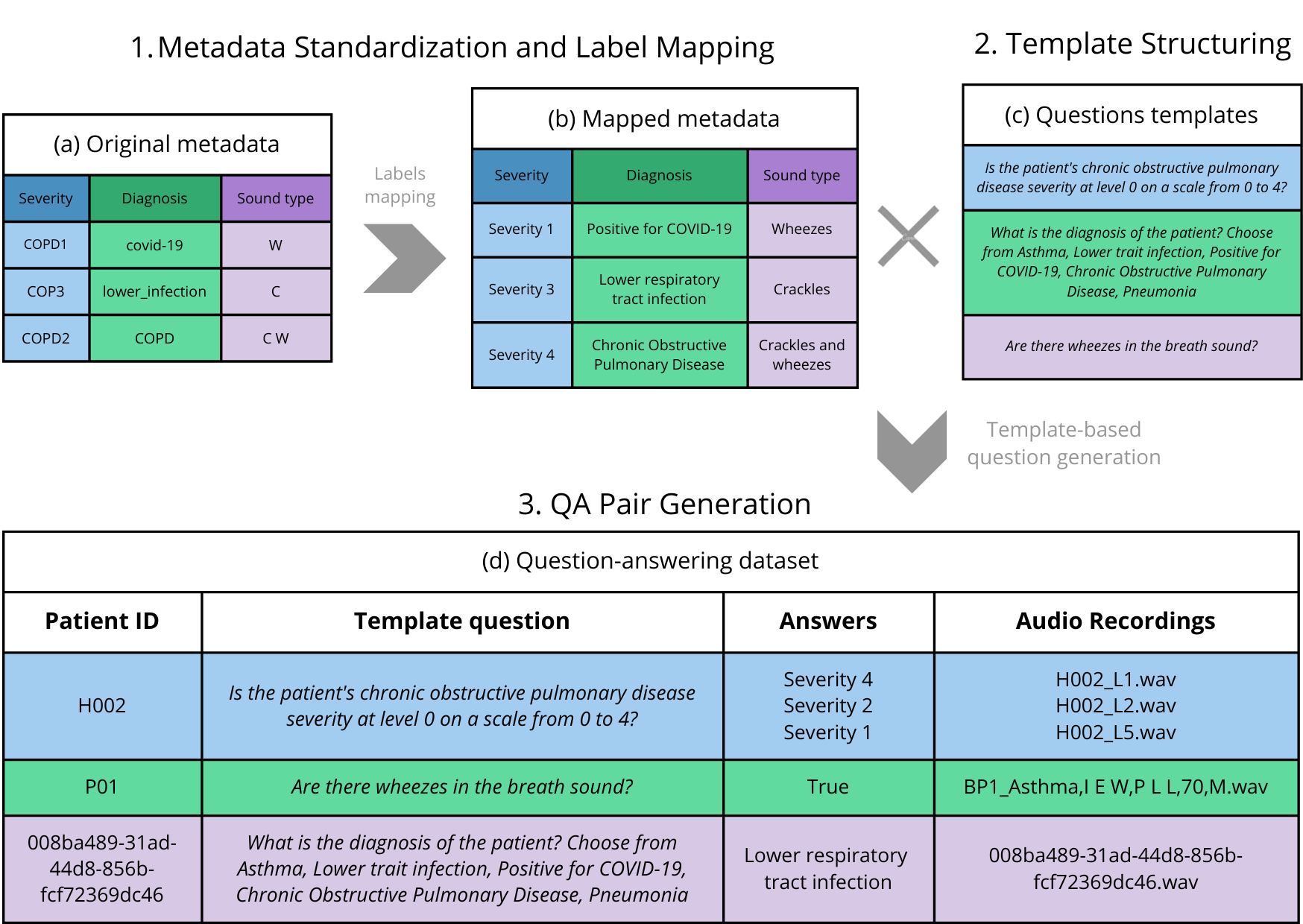}
  \caption{Overview of the RA-QA dataset creation pipeline.}
  \label{fig:pipeline}
\end{figure}

%% file: sections/4_benchmarking.tex
\section{RA-QA Baseline Benchmarking}\label{sec:benchmarking}
RA-QA benchmarking provides reference evaluations for respiratory audio question answering. Our goal is to quantify: (i) what capabilities audio-only models exhibit across the RA-QA task space when relying solely on respiratory acoustics; (ii) how well general-purpose audio-language models transfer to clinically grounded respiratory QA without domain-specific fine-tuning; and (iii) how a state-of-the-art domain-trained audio-language architecture handles the diversity of RA-QA, including heterogeneous respiratory recordings, question types, and answer formats. 

% \yz{make it clear that which are references which are actual benchmarked models and methods? Start with the actual methods, and then follow to say briefly that for understanding the difficulty of the task, we also provide majority guessing, and we provide audio-only methods for understanding the task upper bound without having to cope with the question and task heterogeneity.} \gb{done}
\subsection{Baselines}\label{sec:baselines}

We benchmark models that explicitly address the RA-QA setting by using both the respiratory audio and the question. These include multimodal classifiers, general-purpose audio-language models, and a CaReAQA-style generative baseline. To interpret task difficulty, we also report audio-only models as reference baselines. These baselines are attribute- and question-type-specific, requiring multiple separately trained models to cover the benchmark. They therefore do not solve the full heterogeneous RA-QA setting as a unified question-answering problem, but serve as task-specific references for interpreting difficulty and audio-only capability. 
% \yz{unclear}\gb{done}

\noindent \textbf{Question-conditioned multimodal classifier baselines.}
To study the behaviour of predictive models under explicit question-audio conditioning, we train late-fusion multimodal classifiers using each encoder from OPERA, a state-of-the-art respiratory audio encoding framework~\cite{b6}. We consider its three encoders separately, OPERA-CT, OPERA-GT, and OPERA-CE, to examine how their different training objectives affect the encoding of clinically relevant information for RA-QA. For each OPERA variant, we extract a frozen audio representation and pair it with a fixed text representation of the corresponding question. The resulting audio and question embeddings are then concatenated and passed to an MLP prediction head. We train each classifier over the considered experimental collection, rather than training separate models for each attribute or question type. Categorical samples are handled with a shared classification head over the canonical answer space, while continuous-valued samples are handled with a regression head. This setup evaluates whether a single question-conditioned predictive model can jointly handle multiple attributes, question types, and answer formats.

\noindent \textbf{General audio-language baselines.}
We evaluate two general-purpose audio-language models, namely \textbf{Qwen2.5-Omni}~\cite{b33} and \textbf{Audio Flamingo 3}~\cite{b34}, to assess how far strong pretrained generative systems transfer to clinically grounded respiratory QA without domain-specific fine-tuning. Unlike the attribute-specific classifier baselines, these models are evaluated on the experimental collection as a single evaluation pool, rather than separately by attribute or question type. Each model is prompted with the paired input $(x,q)$ and generates a free-form answer, which is then normalized and parsed for task-level scoring.
% We report each model under two prompting conditions: a question-only setting, where the model receives only the input question, and a task-instructed setting, where the question is preceded by the following respiratory QA instruction: \textit{You are an expert assistant for respiratory audio question answering. Answer the user's question using only the acoustic evidence in the respiratory audio.}}

\noindent \textbf{CaReAQA-style generative baseline.}
We include a domain-trained generative audio-conditioned LLM baseline inspired by CaReAQA, a state-of-the-art generative model for respiratory audio question answering~\cite{b17}. This baseline maps medical-audio embeddings into the language-model embedding space and injects them as a soft audio prefix for decoder-only generation. In contrast to the attribute- or question-type-specific classifier baselines, the CaReAQA-style model is trained and evaluated on the experimental subcollection as a single pool, jointly covering multiple attributes, question types, and answer formats. We encode each recording with a frozen OPERA-CT backbone to obtain audio embeddings and use GPT-2 as the language model.

% \yz{We also report...}

% \noindent \textcolor{red}{\textbf{Majority reference baseline.}
% To quantify the performance obtainable from label priors alone, we include a majority-label predictor that outputs the most frequent training label for each attribute. This reference baseline ignores both the respiratory audio and the question, and therefore does not represent a RA-QA model. Instead, it provides a lower-level reference point for interpreting whether learned models go beyond dataset priors.}

\noindent \textbf{Audio-only reference baselines.}
We also report audio-only reference baselines to estimate the amount of clinically relevant information recoverable from respiratory acoustics alone. Specifically, we train audio-only MLP probes on frozen representations from OPERA encoders. Since the same respiratory recording can be associated with different questions and answer formats, audio-only classifiers are trained separately for each attribute and question type. However, these models do not address the full RA-QA setting, because they do not receive the question text and therefore cannot disambiguate the intended question, target attribute, or answer format. Instead, they provide a task-specific audio-only reference point for each target under a simplified setting.

\input{tables/overview_results}

\subsection{Evaluation Metrics}
\label{sec:benchmark_metrics}
RA-QA answers can be evaluated along two complementary axes: \textit{linguistic correctness}, which measures how closely a predicted answer matches the reference in wording and meaning (i.e., surface form and semantics), and \textit{clinical correctness}, which measures whether the prediction corresponds to the correct underlying clinical label or numeric value, and therefore whether the system is clinically reliable. To capture both aspects, we use two metrics: a \textit{text-level} metric for semantic fidelity and a \textit{task-level} metric for clinical correctness.

\noindent \textbf{Semantic level metric.}
Semantic correctness (text-level fidelity) is intended to capture whether a system produces an answer that is linguistically well-formed and semantically aligned with the reference, even when multiple phrasings are acceptable. In this paper, we report \textbf{BERTScore}~\cite{b30}, highlighting whether predictions preserve the intended meaning beyond exact lexical overlap.

\noindent \textbf{Label-correctness metric.}
Clinical relevance is assessed at the \textit{task level} by extracting the underlying target from the generated answer and treating it as a label. In this paper, we report \textbf{Accuracy} for discriminative tasks. For \textit{regression} targets, clinical correctness is naturally expressed as numeric error, thus we report \textbf{MAE} (mean absolute error), which quantifies the average deviation between predicted and ground-truth values. 

\subsection{Evaluation setup}\label{sec:eval_setup}
We construct a label-balanced subset (with sampling performed \textit{within each attribute}) with a patient-wise split, using $\approx$100{,}000 QA instances for training, $\approx$15{,}000 for validation, and $\approx$15{,}000 for testing. 
% Since a single recording can support multiple questions, for \textit{unimodal} baselines we treat each attribute as an independent classification task, including regressive ones.
% In contrast, \textit{multimodal} baselines are evaluated on the full subset.

%% file: tables/overview_results.tex
\begin{table*}[htbp]
\caption{Baseline performance on RA-QA, grouped by evaluation scope. Asterisks (*) denote models evaluated in a zero-shot setting.}
% , while daggers ($\dagger$) denote prompted models. }
\label{tab:results}
\centering
\scriptsize
\setlength{\tabcolsep}{3.5pt}

\begin{tabular}{lllcc|cc|cc|cc}
\toprule
\multirow{3}{*}{\textbf{Type}}
& \multirow{3}{*}{\textbf{Baseline}}
& \multirow{3}{*}{\textbf{Audio Encoder}}
& \multicolumn{6}{c|}{\textbf{Discriminative}}
& \multicolumn{2}{c}{\textbf{Regressive}} \\
\cmidrule(lr){4-9} \cmidrule(lr){10-11}

&
&
& \multicolumn{2}{c|}{\textbf{Single-verify}}
& \multicolumn{2}{c|}{\textbf{Open-ended}}
& \multicolumn{2}{c|}{\textbf{Multiple-choice}}
& \multicolumn{2}{c}{\textbf{Open-ended}} \\
\cmidrule(lr){4-5} \cmidrule(lr){6-7} \cmidrule(lr){8-9} \cmidrule(lr){10-11}

&
&
& \textbf{Accuracy} & \textbf{BERTScore}
& \textbf{Accuracy} & \textbf{BERTScore}
& \textbf{Accuracy} & \textbf{BERTScore}
& \textbf{MAE} & \textbf{BERTScore} \\
\midrule

\multirow{5}{*}{\textbf{\shortstack{References}}}
& Per-attribute Random
& --
& 0.50 & --
& 0.02 & --
& 0.02 & --
& 6.73 & -- \\

\cmidrule(lr){2-11}

& \multirow{4}{*}{\shortstack{Per-attribute and -question-type\\Audio-only classifier}}
& SVM
& 0.70 & --
& 0.44 & --
& 0.51 & --
& 0.55 & -- \\

& 
& OPERA-CT
& 0.60 & 0.86
& 0.39 & 0.85
& 0.38 & 0.93
& 0.60 & 0.85 \\

&
& OPERA-GT
& 0.57 & 0.86
& 0.40 & 0.89
& 0.40 & 0.89
& 0.61 & 0.85 \\

&
& OPERA-CE
& 0.60 & 0.86
& 0.38 & 0.89
& 0.45 & 0.89
& 0.64 & 0.85 \\

% \cmidrule(lr){2-11}

% & \multirow{3}{*}{Per-attribute multimodal classifier}
% & OPERA-CT
% & 0.64 & 0.86
% & 0.32 & 0.84
% & 0.47 & 0.93
% & 1.09 & 0.85 \\

% &
% & OPERA-GT
% & 0.59 & 0.86
% & 0.37 & 0.84
% & 0.54 & 0.92
% & 1.56 & 0.85 \\

% &
% & OPERA-CE
% & 0.65 & 0.85
% & 0.38 & 0.84
% & 0.46 & 0.93
% & 0.64 & 0.85 \\

\midrule

\multirow{9}{*}{\textbf{\shortstack{Collection-level}}}
% \cmidrule(lr){2-11}

% & Pengi*
% & --
% & -- & --
% & -- & --
% & -- & --
% & -- & -- \

& Audio Flamingo 3*
& --
& 0.44 & 0.92
& 0.02 & 0.89
& 0.37 & 0.93
& 1.26 & 0.88 \\

\cmidrule(lr){2-11}

& Qwen2.5-Omni*
& --
& 0.46 & 0.91
& 0.08 & 0.91
& 0.42 & 0.91
& 1.46 & 0.88 \\

% & Qwen2.5-Omni*$\dagger$
% & --
% & 0.50 & 0.87
% & 0.03 & 0.83
% & 0.14 & 0.94
% & 2.02 & 0.82 \\

% & Audio Flamingo 3*$\dagger$
% & --
% & 0.56 & 0.83
% & 0.00 & 0.85
% & 0.14 & 0.92
% & 9.04 & 0.86 \\

\cmidrule(lr){2-11}

& \multirow{3}{*}{Global multimodal classifier}
& OPERA-CT
& 0.47 & 0.85
& 0.30 & 0.85
& 0.32 & 0.90
& 0.59 & 0.85 \\

&
& OPERA-GT
& 0.41 & 0.85
& 0.29 & 0.85
& 0.32 & 0.90
& 0.59 & 0.85 \\

&
& OPERA-CE
& 0.65 & 0.85
& 0.27 & 0.85
& 0.31 & 0.92
& 0.58 & 0.85 \\

\cmidrule(lr){2-11}

& \multirow{1}{*}{CaReAQA-style}
& -- %OPERA-CT
& 0.52 & 0.94
& 0.22 & 0.97
& 0.51 & 0.96
& 1.91 & 0.99 \\

% &
% & OPERA-GT
% & -- & --
% & -- & --
% & -- & --
% & -- & -- \\

% &
% & OPERA-CE
% & -- & --
% & -- & --
% & -- & --
% & -- & -- \\

\bottomrule
\end{tabular}
\end{table*}

%% file: sections/5_results_old.tex
\section{Results}
\label{sec:results}
% \textbf{Unimodal baselines.} \gb{discussion about the opera + MLP}
% Table~\ref{tab:results} reports random/majority as lower-bound references and shows that the dataset- and task-specific audio-only baseline substantially improves over these priors, especially for open-ended and multiple-choice discriminative questions (MacroF1 $0.49$ and $0.57$), indicating an \textit{informative} respiratory acoustic signal. We note that audio-only is an oracle-style reference that requires training many specialized models (per attribute/dataset/question type), which is operationally unrealistic; we include it as an upper bound on what acoustics alone, without question conditioning, can achieve.
\textbf{Reference baselines.} 
Table~\ref{tab:results} reports random as lower-bound references and shows that the dataset- and task-specific audio-only baseline substantially improves over these priors for both regressive and discriminative questions, indicating an \textit{informative} respiratory acoustic signal. We note that audio-only is an oracle-style reference that requires training many specialized models (per attribute/dataset/question type), which is operationally unrealistic; we include it as a reference on what acoustics alone, without question conditioning, can achieve.
% Audio-only classifiers, despite being trained per attribute and per question type, reveal the inherent difficulty of the benchmark, with average performance barely exceeding — or in some cases failing to surpass — the Majority baseline\yz{is this correct? sounds off to me..They are task and question specific right? Should not be random?} \yz{why is this different from last submission?}. The per-attribute multimodal classifier yields consistent improvements across all tasks, with Macro F1 reaching up to 0.57 on Single-verify and 0.37 on Multiple-choice, and MAE dropping to below 0.65 on the regressive task compared to 412.65 for the Majority baseline. However, this multimodal approach remains inherently fragmented: by construction, it cannot capture cross-attribute dependencies, reason jointly across task types, or produce contextually coherent responses. 

\noindent \textbf{General audio-language transfer.}
The general audio-language baselines, Qwen2.5-Omni and Audio Flamingo 3, show limited and uneven transfer to RA-QA (Table~\ref{tab:results}). Their task-level accuracy is particularly low for open-ended discriminative questions, reaching only 0.08 for Qwen2.5-Omni and 0.02 for Audio Flamingo 3. Performance is higher on multiple-choice questions, reflecting the better instruction-following behaviour of these audio-language models. At the same time, their BERTScore values remain relatively high across settings, suggesting that lexical or semantic overlap metrics can overestimate answer quality when models generate fluent but generic responses. Qualitatively, both models often default to broad audio descriptions or plausible answers rather than clinical-sounding predictions. This behaviour highlights the need for RA-QA models and benchmarks tailored to clinically grounded respiratory audio reasoning.

 \noindent \textbf{Trained multimodal baselines.}
% Both the multimodal classifier and the CaReAQA-style generative baseline were trained jointly on the pooled RA-QA subset, mixing datasets, attributes, and question types. Compared to Pengi, the multimodal classifier yields markedly higher discriminative MacroF1 (0.59 on single-verify), showing that question conditioning helps disambiguate intent under heterogeneous data; however, it is non-generative and thus limited to predicting canonical targets rather than producing free-form answers. 
% The CaReAQA-style model improves over zero-shot transfer and attains a strong BERTScore (up to 0.96), indicating that respiratory-trained audio-to-LLM alignment is beneficial, but task-level performance remains modest on open-ended and multiple-choice questions. Overall, these results underline the benchmark's core challenge of handling respiratory audio question answering heterogeneity and motivates for innovative approaches.
 Both the multimodal classifier and the CaReAQA-style generative baseline were trained jointly on the pooled RA-QA subset, mixing datasets, attributes, and question types. Compared with the best zero-shot general audio-language baseline for each column, the best global multimodal classifier improves single-verify accuracy from 0.46 to 0.65 (+0.19) and open-ended discriminative accuracy from 0.08 to 0.30 (+0.22), while reducing regression MAE from 1.26 to 0.58 ($-$0.68); however, it is non-generative and thus limited to predicting canonical targets rather than producing free-form answers. The CaReAQA-style model also improves over zero-shot transfer on all discriminative question types, increasing accuracy from 0.46 to 0.52 on single-verify questions (+0.06), from 0.08 to 0.22 on open-ended discriminative questions (+0.14), and from 0.42 to 0.51 on multiple-choice questions (+0.09). It also attains the strongest BERTScore values (up to 0.99 for regressive tasks), indicating that respiratory-trained audio-language alignment is beneficial. Overall, these results underline the benchmark's core challenge: handling RA-QA's heterogeneity requires models that jointly optimize semantic fidelity \textit{and} task-level correctness.

\noindent \textbf{Case studies.}
To further analyze benchmark behavior beyond aggregate scores, we report two representative cases in Table~\ref{tab:usecases_results}. In case \textbf{(1)}, we evaluate ICBHI \textit{diagnosis} tasks across different question formats. The CaReAQA-style model obtains the best task-level accuracy across \texttt{single-verify}, \texttt{open-ended}, and \texttt{multiple-choice} questions, but its performance varies substantially by format, decreasing from 0.84 on \texttt{single-verify} to 0.28 on \texttt{open-ended} and 0.59 on \texttt{multiple-choice}. Moreover, task-level accuracy and semantic fidelity do not always align: for example, on open-ended questions, the multimodal classifier obtains higher BERTScore than single-verifies despite failing to recover the correct label. This highlights the need to evaluate both task-level correctness and text-level semantic similarity when assessing generative respiratory QA systems.
In case \textbf{(2)}, we analyze \textit{exhalation SNR} from the self-reported UK COVID-19 dataset. Among trained baselines, the CaReAQA-style model is competitive with the multimodal regression baseline in terms of MAE (0.08 vs.\ 0.06), while also producing natural-language outputs with high semantic fidelity (BERTScore 0.94). This suggests that question-conditioned generative models can remain competitive on continuous-valued respiratory attributes while retaining the advantage of natural-language answer generation.

% \noindent \textbf{Case studies.} To further analyze benchmark behavior beyond aggregate scores, we report two representative cases summarized in Table~\ref{tab:usecases_results}. For case \textbf{(1)}, we report results on ICBHI \textit{diagnosis} tasks, which show that baseline ranking depends strongly on question format: the CaReAQA-style model performs best on \texttt{single-verify}, \texttt{open-ended}, and \texttt{multiple-choice}. This suggests that generative audio-to-LLM pipelines take advantage from flexible question generation. For case \textbf{(2)}, we analyze \textit{exhalation SNR} from the self-reported dataset UK COVID-19. Among trained baselines, the CaReAQA-style model is competitive with the multimodal classifier on the regressive attribute (MAE 0.08 vs.\ 0.06) while maintaining high semantic fidelity (BERTScore 0.94). This suggests that question-conditioned generation can be robust to self-recorded variability, motivating stronger generative baselines for continuous targets under RA-QA heterogeneity, which can remain competitive with classification methods while retaining the advantage of natural-language expression.
\input{tables/usecases_results}

\noindent \textbf{Performance across audio types.} The in-domain CareAQA-style baseline exhibits slightly different performance patterns across audio types.
 Breath achieved the highest accuracy (0.5536), followed by speech (0.5448), while cough was slightly lower (0.5295). Among speech-like recordings, counting outperformed sustained vowels (0.4118 vs 0.3769), where /o/ achieved the highest accuracy (0.3856) and /e/ was the weakest (0.3660), and normal counting was slightly stronger than fast counting (0.4183 vs 0.4052). These findings suggest that audio modality influences respiratory-health QA performance, with breath and speech achieving higher accuracy than cough, and counting outperforming sustained phonation.

\noindent \textbf{Performance across question format.}
The results show a consistent hierarchy of difficulty across question types. Single-verify questions are generally the easiest, reflecting their simpler verification structure. Multiple-choice questions are more challenging, but the presence of explicit candidate answers still provides useful guidance to the model. In contrast, open-ended discriminative questions are the most difficult, as they require the model to infer the target attribute and produce the correct answer without being guided by predefined options. This pattern shows that in respiratory audio question-answering difficulty is not determined only by the underlying clinical attribute, but also by the interaction format and the degree of constraint imposed on the answer space.

% \noindent \textbf{BERTScore sensitivity.}
% Across all tasks and baselines, BERTScore remains consistently high even in cases where the corresponding task-level metric (Macro F1 or MAE) reveals poor or near-random performance. This indicates that models and baselines tend to produce linguistically fluent and semantically plausible outputs, even when their answers are factually incorrect or clinically inconsistent. While linguistic fluency is a desirable property, it is insufficient in clinical contexts where correctness and reliability are paramount: a response can be well-formed and semantically coherent yet convey the wrong diagnosis, severity level, or attribute value. This result motivates the decoupling between surface-level language quality and task-level accuracy motivates reporting BERTScore alongside, in order to ensure evaluation reflects clinical relevance along with linguistic adequacy.}

%% file: tables/usecases_results.tex
\begin{table}[htbp]
\centering
\scriptsize
\caption{Case-studies results by question type (SV / OE / MC). Each cell reports task score (Accuracy or MAE) / BERTScore.}
\label{tab:usecases_results}
\begin{tabular}{llccc}
\toprule
\textbf{Case} & \textbf{QT} & \textbf{Multimodal classifier} & \textbf{Qwen2.5-Omni} & \textbf{CareAQA-style} \\
\midrule
\multirow{3}{*}{\textbf{Case 1}}
 & SV & 0.34 / 0.82 & 0.05 / 0.82 & 0.84 / 0.96 \\
 & OE & 0.00 / 0.87 & 0.09 / 0.87 & 0.28 / 0.67 \\
 & MC & 0.01 / 0.92 & 0.00 / 0.83 & 0.59 / 0.83 \\
\midrule
\textbf{Case 2}
 & OE & 0.06 / 0.85 & 1.33 / 0.85 & 0.08 / 0.94 \\
\bottomrule
\end{tabular}
\end{table}

%% file: sections/6_conclusions.tex
\section{Conclusion}
\label{sec:conclusion}
We introduced \textbf{RA-QA}, an open-source \textit{collection and benchmarking system} for respiratory audio question answering, comprising a standardized data-curation pipeline, leakage-aware splits, and a unified evaluation protocol. RA-QA is designed to stress-test models under realistic heterogeneity, including multiple question formats over the same recording; diverse modalities and datasets/devices; and both discriminative and regression targets across four attribute families. Baseline results show that RA-QA's heterogeneity exposes the limited robustness of current approaches and that high semantic similarity can still correspond to low task-level clinical correctness, motivating heterogeneity-aware respiratory-audio QA models that jointly optimize linguistic fidelity and prediction accuracy. By releasing RA-QA as a reproducible benchmarking system, we aim to enable fair comparisons under heterogeneity, and to support the development of respiratory-specific, question-conditioned models that are both linguistically well-formed and task-correct.

%% file: appendix/datasets.tex
\section{Datasets overview}\label{apd:datasets_overview}

RA-QA collection is constructed from 11 distinct source datasets, each contributing complementary coverage of recording modalities, acquisition conditions, and clinical labels. This appendix summarizes these datasets in more detail. For each source, we rely on two primary elements: (i) a metadata file containing structured clinical and contextual annotations, and (ii) the associated respiratory audio recordings. All recordings are anonymized, and metadata are curated to remove personally identifiable information, in line with ethical and data protection requirements. This consistent organization simplifies integration across sources and supports reproducible downstream experiments. We next describe each dataset included in RA-QA.
\\ \\
\textbf{COVID-19 Sounds} \cite{b18}
COVID-19 Sounds contains 53{,}449 audio samples (over 552 hours) collected from 36{,}116 participants through the COVID-19 Sounds mobile application. The cohort spans diverse demographics and health conditions, including 2{,}106 samples from participants who self-reported as COVID-19 positive (see Figure~\ref{fig:covid19_summary_plot}). The dataset includes three audio modalities: breathing, cough, and voice. Data collection received approval from the Ethics Committee of the Department of Computer Science and Technology at the University of Cambridge, and informed consent was obtained from all participants. Access is granted under a Data Transfer Agreement, and the dataset has been widely used in prior work \cite{b35, b36}.
\\ \\
\textbf{UK COVID-19} \cite{b19}
The UK COVID-19 Vocal Audio Dataset was created to train and evaluate machine learning models for SARS-CoV-2 infection status and related symptoms from vocal audio. It was collected by the UK Health Security Agency between March 2021 and March 2022, covering periods dominated by the Alpha and Delta variants and including some Omicron sublineages. Participants were recruited via the national Test and Trace programme and the REACT-1 survey. Voluntary cough and exhalation recordings were gathered through the ``Speak up to help beat coronavirus'' digital survey, together with demographics, self-reported symptoms, and respiratory conditions (see Figure~\ref{fig:uk_covid_19_summary_plot}). These recordings were linked to SARS-CoV-2 test results; speech recordings are not part of the open-access release and are not used in this study. The study received approval from The National Statistician’s Data Ethics Advisory Committee (NSDEC(21)01), the Cambridge South NHS REC (21/EE/0036), and the Nottingham NHS REC (21/EM/0067), with informed consent obtained from all participants.
\\ \\
\textbf{COUGHVID} \cite{b20}
COUGHVID includes more than 25{,}000 crowdsourced cough recordings contributed by participants across a broad range of ages, genders, locations, and COVID-19 statuses (see Figure~\ref{fig:coughvid_summary_plot}). Collection and annotation followed relevant ethical guidelines, and participants provided consent to share their cough recordings and associated metadata.
\\ \\
\textbf{ICBHI} \cite{b21}
The ICBHI Respiratory Sound Database aggregates recordings collected by two independent teams over multiple years and countries. Most samples were recorded by the School of Health Sciences, University of Aveiro (ESSUA) group at Lab3R in Aveiro, Portugal and Hospital Infante D. Pedro. A second collection effort from Aristotle University of Thessaloniki (AUTH) and the University of Coimbra (UC) recorded at Papanikolaou General Hospital (Thessaloniki) and the General Hospital of Imathia (Naousa), Greece. Ethical approval was obtained from the relevant institutional committees. The dataset contains 5.5 hours of audio across 920 annotated recordings from 126 subjects (see Figure~\ref{fig:icbhi_summary_plot}).
\\ \\
\textbf{HF Lung} \cite{b22}
HF Lung V2 comprises two subsets, HF Lung V1 and HF Lung V1 IP. HF Lung V1 combines (i) recordings collected during a 2020 datathon organized by the Taiwan Smart Emergency and Critical Care (TSECC), released under CC BY 4.0 by the Taiwan Society of Emergency and Critical Care Medicine (TSECCM), including 261 patients, and (ii) recordings from 18 patients in a respiratory care ward/center in Northern Taiwan (Aug 2018--Oct 2019), approved by the Research Ethics Review Committee of Far Eastern Memorial Hospital (case 107052-F) with written informed consent. HF Lung V1 IP similarly includes (i) 32 patients recorded with a Littmann 3200 digital stethoscope provided by TSECCM and (ii) 7 residents from a respiratory care ward/center in Northern Taiwan (Aug--Dec 2019), also approved under case 107052-F with written informed consent from patients or statutory agents. The dataset provides detailed event-level annotations for crackles, wheezes, and stridor (see Figure~\ref{fig:hf_lung_summary_plot}).
\\ \\
\textbf{MM Lung} \cite{b23}
MMLung was collected from 40 UK participants (20 male, 20 female), aged 18--85, all English speakers. Twelve participants were healthy; the remainder included seven self-reported COPD, seven self-reported asthma, and 14 with other long-term conditions. Ethical approval was granted by the University of Southampton. Data were collected using three devices, including a Google Pixel 6 smartphone (custom acquisition app) and an Easy on-PC ultrasonic spirometer (ndd Medical Technologies). Audio was recorded in stereo at 44.1~kHz as WAV in a quiet room. The protocol covers cough, vowels, mobile spirometry, and speech tasks; in this work we use only the deep breath and the vowel ``o'' recordings (see Figure~\ref{fig:mm_lung_summary_plot}). Clinical reference measurements were obtained via a medical-grade spirometer following ATS/ERS standards, noting that effort-dependent tests may introduce variability. The dataset is available upon request.
\\ \\
\textbf{Coswara} \cite{b24}
Coswara contains respiratory audio collected between April 2020 and February 2022 from 2{,}635 individuals: 1{,}819 SARS-CoV-2 negative, 674 positive, and 142 recovered. It includes nine respiratory sound categories spanning breathing, cough, and speech. Metadata include demographics (e.g., age, gender, location) and health information such as symptoms, pre-existing respiratory conditions, comorbidities, and SARS-CoV-2 test status (see Figure~\ref{fig:coswara_summary_plot}). Collection was approved by the Institutional Human Ethics Committee at the Indian Institute of Science, Bangalore; participants provided informed consent and the released data are anonymized.
\\ \\
\textbf{KAUH} \cite{b26}
KAUH provides lung sound recordings from patients with seven respiratory conditions (asthma, heart failure, pneumonia, bronchitis, pleural effusion, lung fibrosis, COPD) as well as normal breathing (see Figure~\ref{fig:KAUH_summary_plot}). Recordings were acquired using an electronic stethoscope at multiple chest-wall locations by a specialist physician, with each sound captured three times using different frequency filters. Written informed consent was obtained from participants (or parents/guardians for minors). The study was approved by the Institutional Review Board at King Abdullah University Hospital and Jordan University of Science and Technology (Ref.~91/136/2020), and the authors report the right to share the data publicly.
\\ \\
\textbf{SSBPR} \cite{b25}
SSBPR is a sleep body-position recognition dataset based on snoring audio, comprising 7{,}570 samples labeled with six sleep positions: supine, supine with left lateral head, supine with right lateral head, left-side lying, right-side lying, and prone (see Figure~\ref{fig:SSBPR_summary_plot}). 
% One label that appears in only a few subjects is excluded, following the 5-class setup in \cite[Sec.~\dots]{xiao2023snoringsounddatasetbody}. 
Data were collected from 20 adults undergoing overnight polysomnography at a local Sleep Medicine Research Center. The study received local ethics approval and participants provided signed consent for audio/video recording; personal information was anonymized.
\\ \\
\textbf{NoseMic} \cite{b27}
NoseMic is derived from a respiratory rate estimation study. Audio was captured using microphones placed near the nose, while respiratory dynamics were measured with a Zephyr chest pressure sensor. Data were recorded in a stationary setting before and after physical exercise. The study includes 21 participants, with some excluded due to poor sensing quality. Recordings are segmented into 30-second windows with 15-second overlap; the average respiratory rate per window serves as ground truth.
\\ \\
\textbf{Respiratory@TR} \cite{b28}
Respiratory@TR contains lung sounds recorded with two digital stethoscopes at Antakya State Hospital, covering left/right posterior and anterior chest, as well as back recordings. In addition to lung sounds, it includes chest X-rays, pulmonary function test (PFT) variables, spirometric curves, and SGRQ-C questionnaire scores. Lung sounds are captured across 12 channels targeting upper/middle/lower regions and costophrenic angles. Recordings were validated and labeled by two pulmonologists using chest X-rays, PFT outcomes, and auscultation, with labels corresponding to five COPD severity levels (COPD0--COPD4). The dataset was released by Iskenderun Technical University, Turkey; participation was voluntary and the cohort spans diverse backgrounds and genders.
\\ \\
Two additional plots summarize demographic variability across datasets by showing the distributions of age and gender (Figure~\ref{fig:age}, ~\ref{fig:gender}).
\input{appendix/figures/covid_19}
\input{appendix/figures/uk_covid_19}
\input{appendix/figures/coughvid}
\input{appendix/figures/icbhi}
\input{appendix/figures/hf_lung}
\input{appendix/figures/mm_lung}
\input{appendix/figures/coswara}
\input{appendix/figures/kauh}
\input{appendix/figures/ssbpr}
\input{appendix/figures/nosemic}
\input{appendix/figures/respiratoryTR}
\input{appendix/figures/age}

\input{appendix/figures/gender}

%% file: appendix/figures/covid_19.tex
\begin{figure*}[p]
  \centering
  \includegraphics[width=1\linewidth]{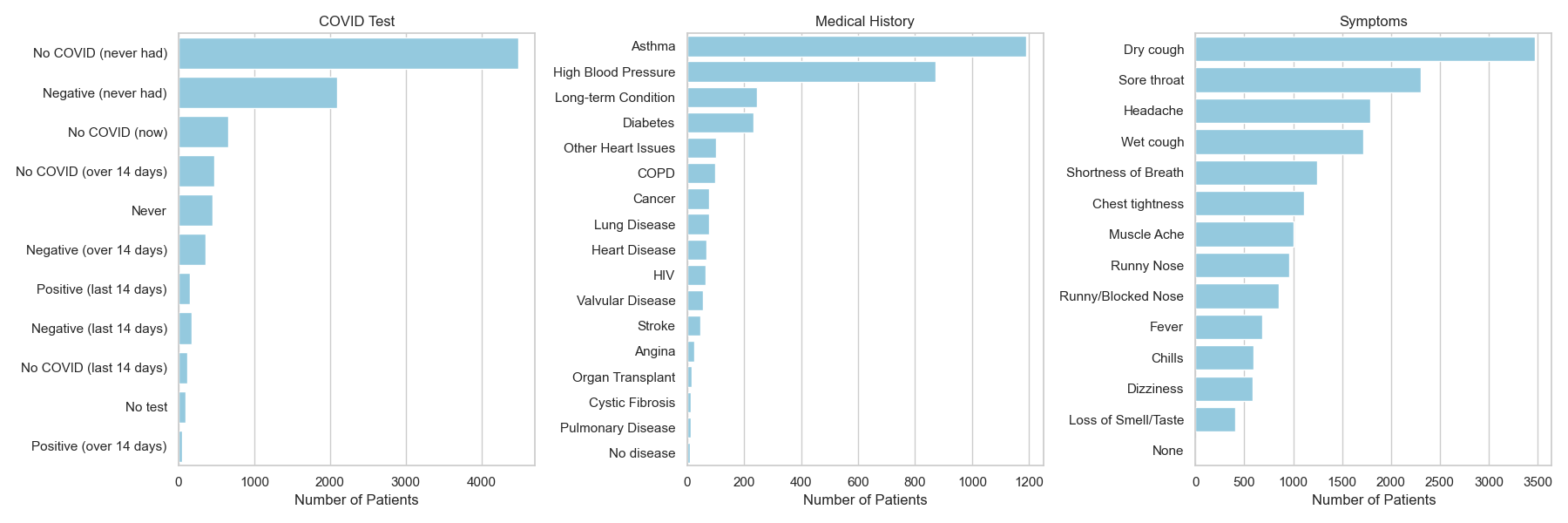}
    \caption{Example metadata from the COVID-19 Sounds dataset. For each respiratory audio recording, the dataset includes the result of a COVID-19 test, along with the corresponding period of positivity or negativity in some cases, as well as self-reported medical history and symptoms. Each individual may report multiple coexisting medical conditions and several symptoms, offering a detailed clinical profile for each recording.}
  \label{fig:covid19_summary_plot}
\end{figure*}

%% file: appendix/figures/uk_covid_19.tex
\begin{figure*}[p]
  \centering
  \includegraphics[width=1\linewidth]{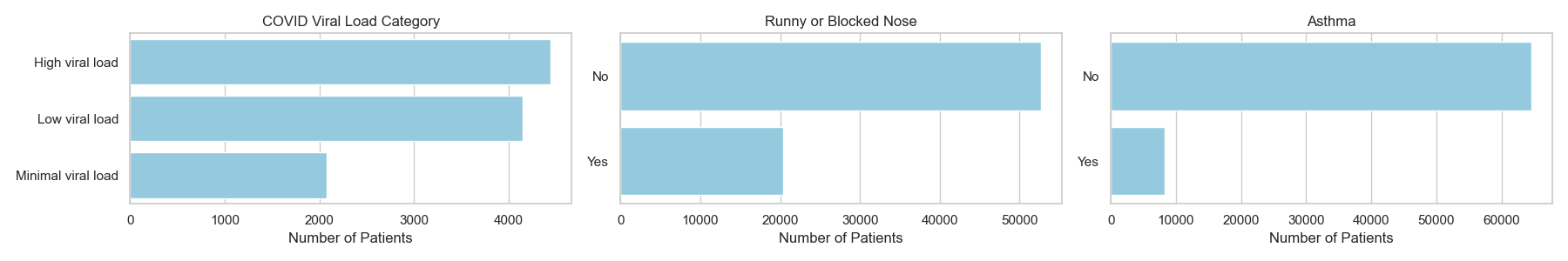}
    \caption{The figure shows label distributions for viral load categories (multiclass), as well as binary labels for symptoms such as runny or blocked nose and conditions like asthma from the UK COVID-19 dataset. As with other datasets used in this work, these labels are highly imbalanced and require preprocessing and reduction strategies to ensure meaningful training and evaluation.}
  \label{fig:uk_covid_19_summary_plot}
\end{figure*}

%% file: appendix/figures/coughvid.tex
\begin{figure*}[p]
  \centering
  \includegraphics[width=1\linewidth]{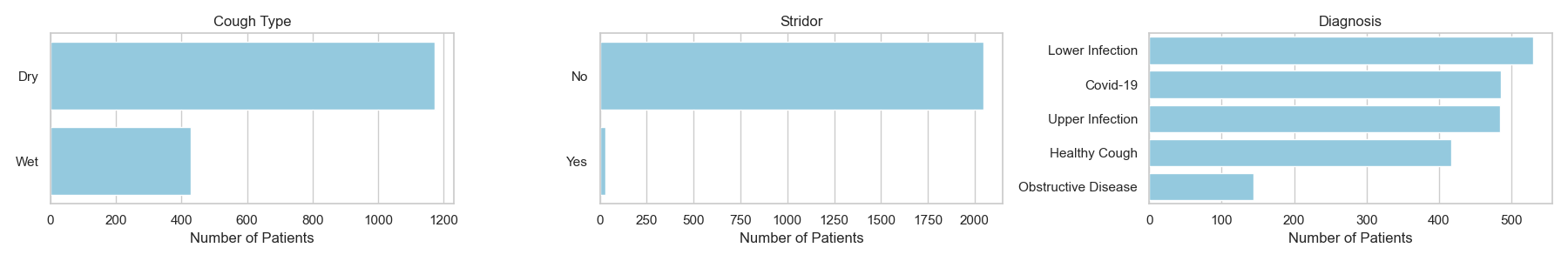}
    \caption{Label distributions from the CoughVid dataset, illustrating examples of audio-related attributes such as cough type, presence of stridor and associated diagnoses. This highlights that, beyond clinical metadata, some datasets also include perceptual or acoustic labels (e.g., wheezes, stridors), which are directly linked to the audio signal and can support more fine-grained sound analysis.}
  \label{fig:coughvid_summary_plot}
\end{figure*}

%% file: appendix/figures/icbhi.tex
\begin{figure*}[p]
  \centering
  \includegraphics[width=1\linewidth]{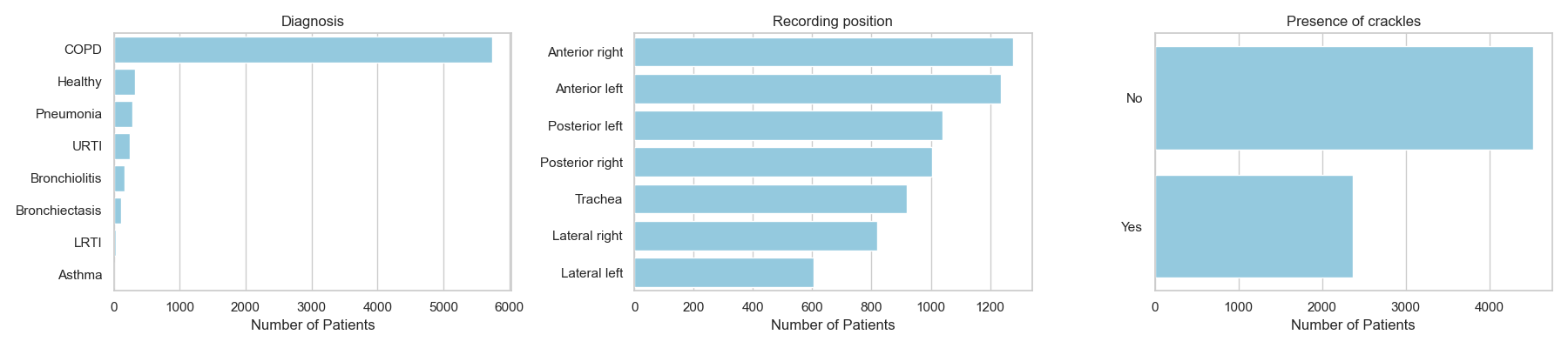}
    \caption{Label distributions from the ICBHI dataset, showing the annotated diagnosis categories, presence of crackles and recording positions (e.g., trachea, anterior left, posterior right). This exemplifies how some datasets provide detailed contextual metadata, such as auscultation position, which can be crucial for interpreting respiratory sounds and modeling location-sensitive acoustic features.}
  \label{fig:icbhi_summary_plot}
\end{figure*}

%% file: appendix/figures/hf_lung.tex
\begin{figure*}[p]
  \centering
  \includegraphics[width=0.4\linewidth]{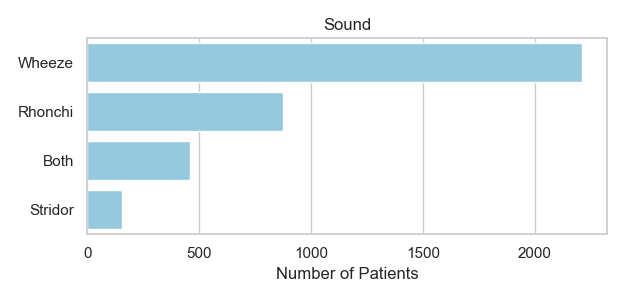}
    \caption{The figure illustrates the distribution of annotated respiratory sound events in the HF Lung dataset, highlighting the presence of clinically relevant audio features such as crackles, wheezes and stridor, providing fine-grained insight into abnormal lung sounds captured in the recordings.}
  \label{fig:hf_lung_summary_plot}
\end{figure*}

%% file: appendix/figures/mm_lung.tex
\begin{figure*}[p]
  \centering
  \includegraphics[width=1\linewidth]{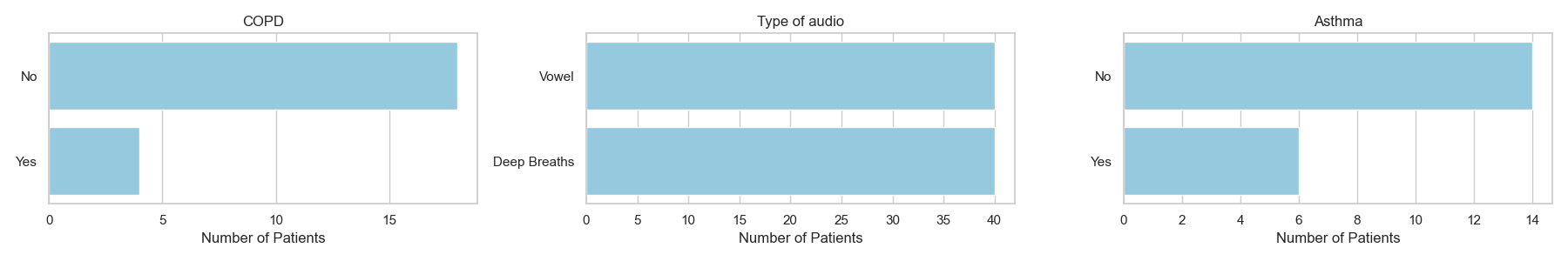}
    \caption{The figure illustrates key characteristics from the MM Lung dataset. The Type of Audio distinguishes between different types of respiratory recordings, such as vowel phonation and deep breathing, which capture varied aspects of lung sound dynamics.}
  \label{fig:mm_lung_summary_plot}
\end{figure*}

%% file: appendix/figures/coswara.tex
\begin{figure*}[p]
  \centering
  \includegraphics[width=1\linewidth]{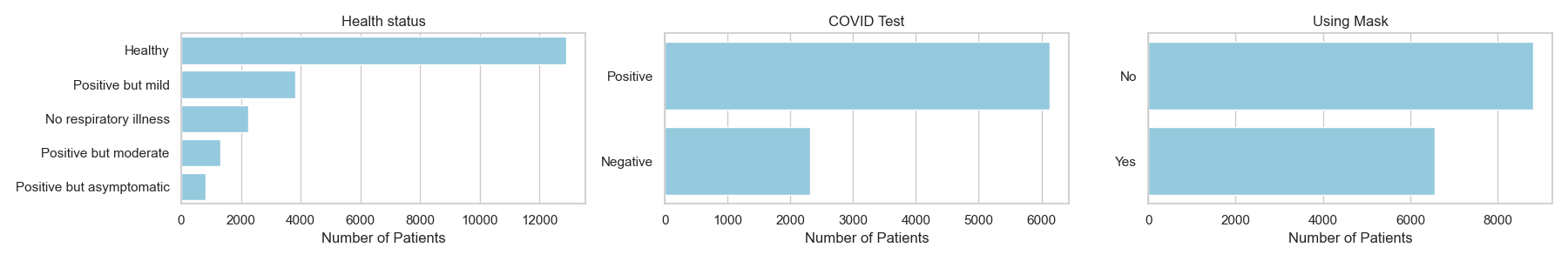}
    \caption{The figure presents the distribution from the Coswara dataset of the selected attributes health status, COVID test result and mask usage at the time of recording. The Health Status attribute distinguishes between healthy individuals and COVID-19 positive cases with varying symptom severity, categorized as asymptomatic, mild or moderate.}
  \label{fig:coswara_summary_plot}
\end{figure*}

%% file: appendix/figures/kauh.tex
\begin{figure*}[p]
  \centering
  \includegraphics[width=1\linewidth]{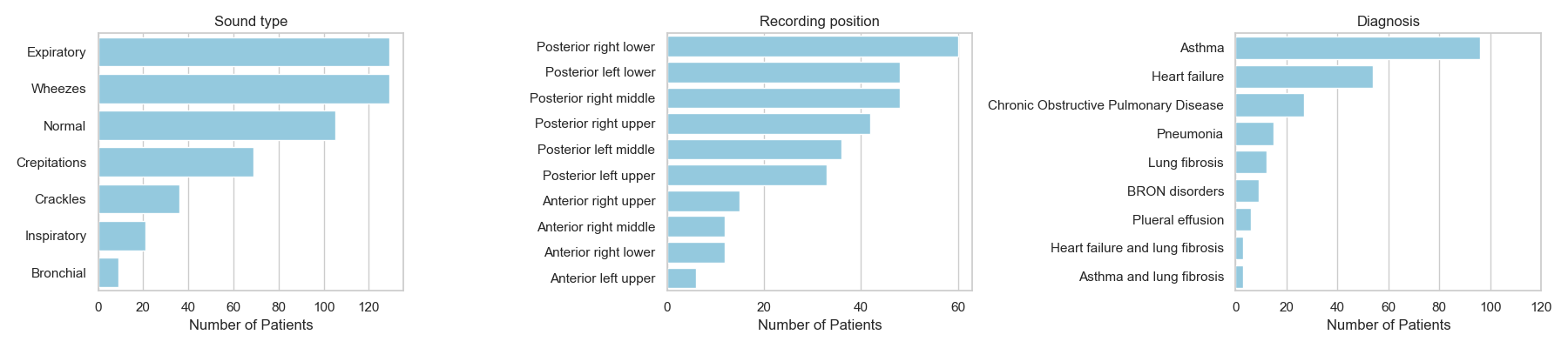}
    \caption{The figure displays key attributes of the KAUH dataset: sound type, recording position and diagnosis. The Sound Type plot shows the presence of multiple annotated labels per recording.}
  \label{fig:KAUH_summary_plot}
\end{figure*}

%% file: appendix/figures/ssbpr.tex
\begin{figure*}[htbp]
  \centering
  \includegraphics[width=0.5\linewidth]{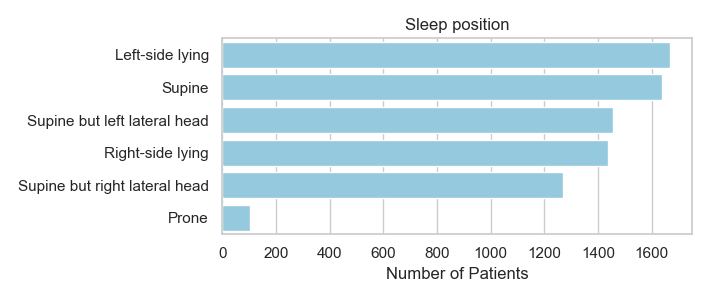}
    \caption{The figure presents the distribution of sleep positions recorded in the SSBPR dataset. Each recording is annotated with the body position of the subject during sleep (e.g., supine, lateral), which can influence the nature and intensity of respiratory sounds.}
  \label{fig:SSBPR_summary_plot}
\end{figure*}

%% file: appendix/figures/nosemic.tex
\begin{figure*}[htbp]
  \centering
  \includegraphics[width=0.4\linewidth]{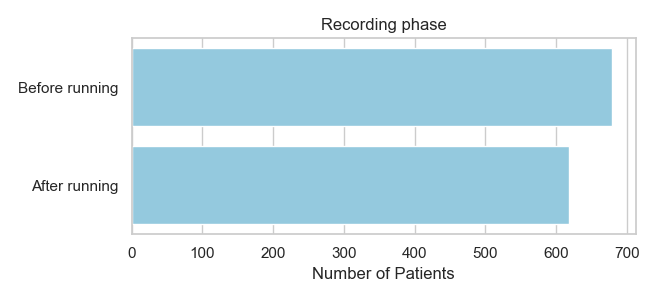}
    \caption{The plots illustrate the recording phase attribute distribution form the NoseMic dataset. This captures whether the respiratory sounds were recorded before or after physical exertion (e.g., running), which is important for evaluating exertion-induced changes in breathing patterns.}
  \label{fig:nosemic_summary_plot}
\end{figure*}

%% file: appendix/figures/respiratoryTR.tex
\begin{figure*}[htbp]
  \centering
  \includegraphics[width=1\linewidth]{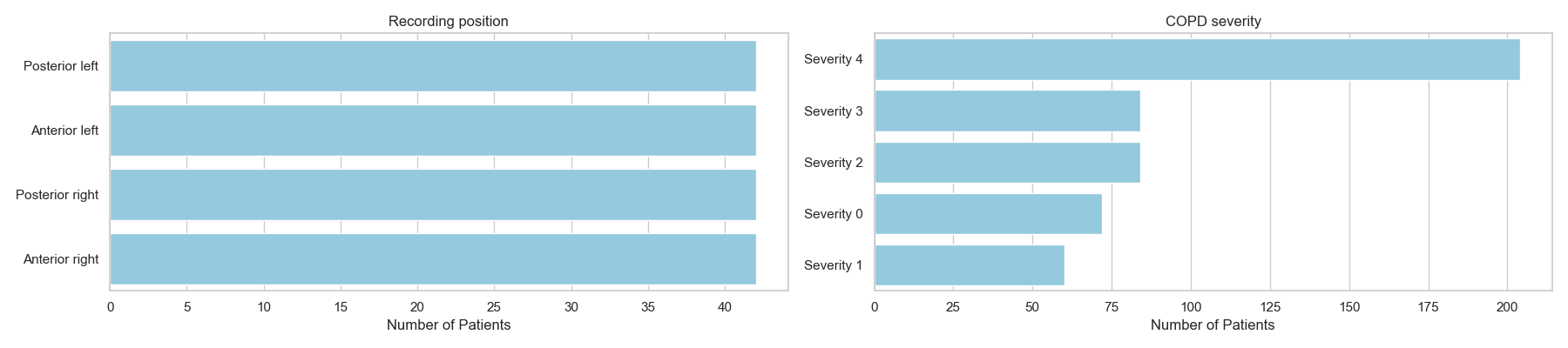}
    \caption{The plots illustrate two key attributes from the Respiratory@TR dataset: recording position and COPD severity. The recording position captures the recording point while the COPD Severity is scored on a scale from 0 (no COPD) to 4 (very severe), providing clinically relevant gradation to analyze how disease progression correlates with acoustic features in the recordings.}
  \label{fig:respiratoryTR_summary_plot}
\end{figure*}

%% file: appendix/figures/age.tex
\begin{figure*}[p]
  \centering
  \includegraphics[width=0.8\linewidth]{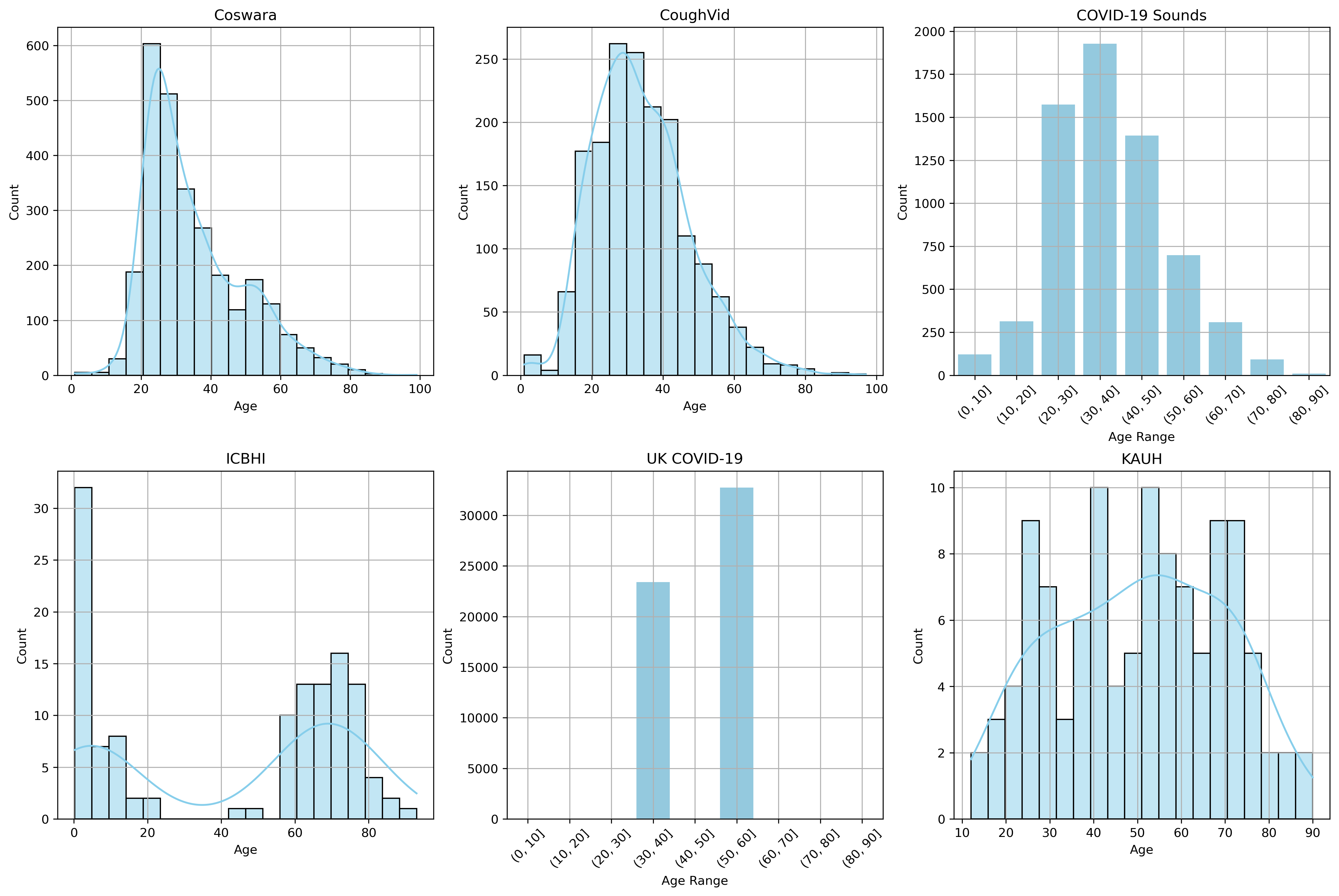}
    \caption{Age distribution distribution for datasets where the information was available.}
  \label{fig:age}
\end{figure*}

%% file: appendix/figures/gender.tex
\begin{figure*}[p]
  \centering
  \includegraphics[width=0.8\linewidth]{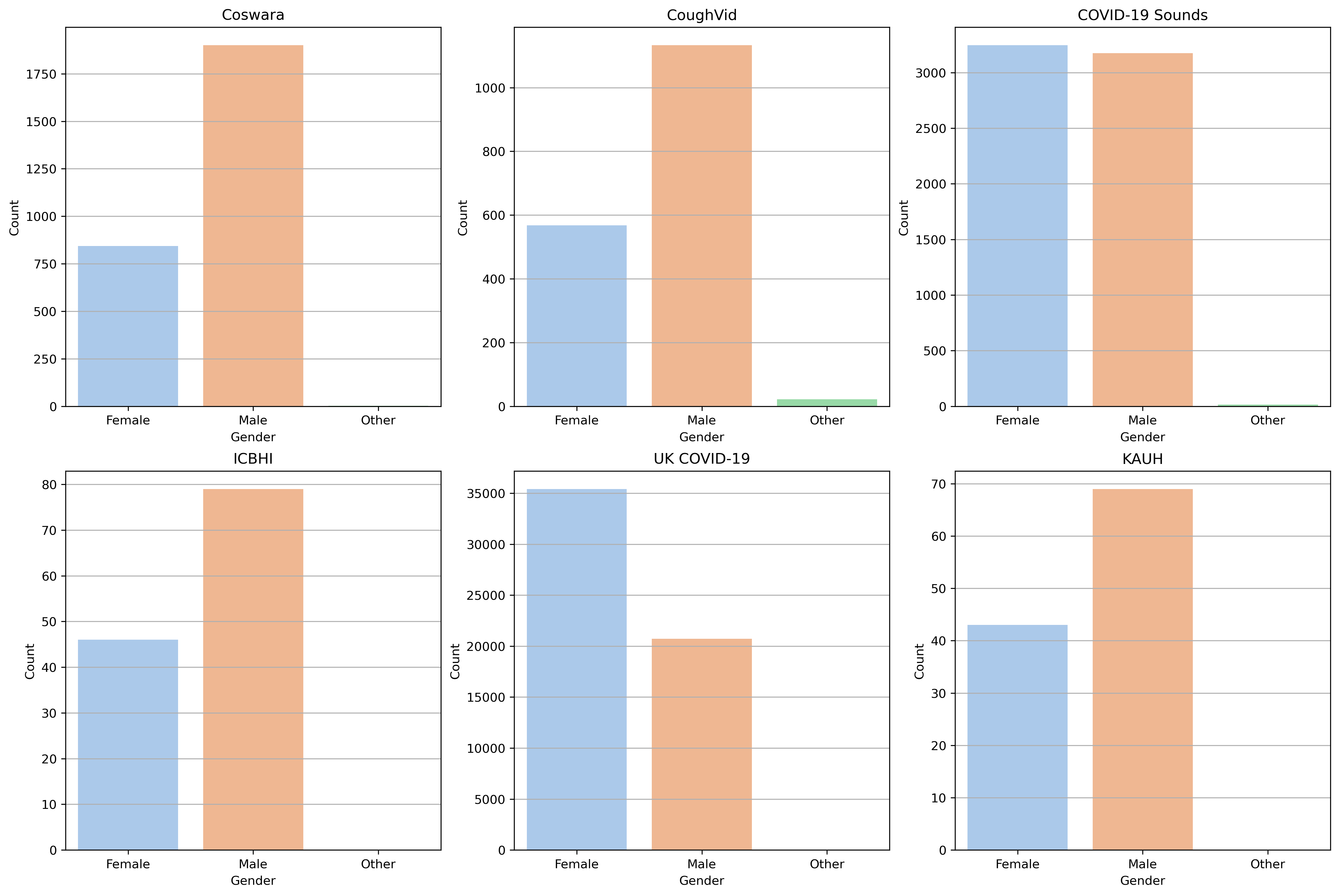}
    \caption{Gender distribution for datasets where the information was available.}
  \label{fig:gender}
\end{figure*}

%% file: appendix/construction.tex
\section{RA-QA Construction Artifacts}\label{apd:construction}
\input{figures/sankey}

This appendix provides additional artifacts that support reproducibility and help the reader interpret the RA-QA construction choices described in the main paper. \\
% RA-QA is derived by harmonizing heterogeneous respiratory datasets into a unified attribute space and then instantiating multiple question formats per attribute. The materials below illustrate the key intermediate representations used by the pipeline and the resulting distribution of QA instances.
\\\noindent \textbf{Label harmonization example.}
RA-QA standardizes dataset-specific labels into consistent, human-readable strings that can be used both as targets for evaluation and as answer options in multiple-choice questions. Table~\ref{tab:mapping_example} shows a concrete mapping example for the \textit{Diagnosis} attribute in \textit{CoughVID}, highlighting how raw annotation values are converted into canonical text that is shared across RA-QA.\\

\noindent \textbf{Question template examples.} To cover realistic interaction styles, RA-QA supports three question formats: open-ended, multiple-choice, and single-verify. Table~\ref{tab:example_of_questions} reports representative templates for multiple attributes across formats, illustrating how the pipeline preserves attribute semantics while varying question intent and answer structure. These templates are used consistently across datasets whenever the underlying attribute is available.\\ 

\noindent \textbf{JSON template specification.} RA-QA question generation is driven by lightweight JSON templates that define the question text, its format, and (when applicable) the set of covered labels. Figure~\ref{lst:json_template_example} shows an example template file containing one open-ended prompt, one multiple-choice prompt with explicit answer options, and one single-verify question derived from a specific label. This format enables regeneration of the dataset and supports future extensions (e.g., adding new paraphrases or attributes) without changing the core pipeline.\\ 

\noindent \textbf{Dataset-level data flow and scale.}
Finally, Figure~\ref{fig:sankey} summarizes the RA-QA construction at the dataset level, visualizing how recordings from each source dataset translate into generated QA instances. The Sankey diagram reports both the number of recordings and the resulting number of questions (log-scaled), making explicit the heterogeneous contribution of different datasets and the effect of generating multiple questions per recording when annotations permit.
\input{appendix/tables/mapping_example}
\input{appendix/tables/example_of_questions}
\input{appendix/tables/json_template}

%% file: figures/sankey.tex
\begin{figure*}[htbp]
  \centering
  \includegraphics[width=1\linewidth]{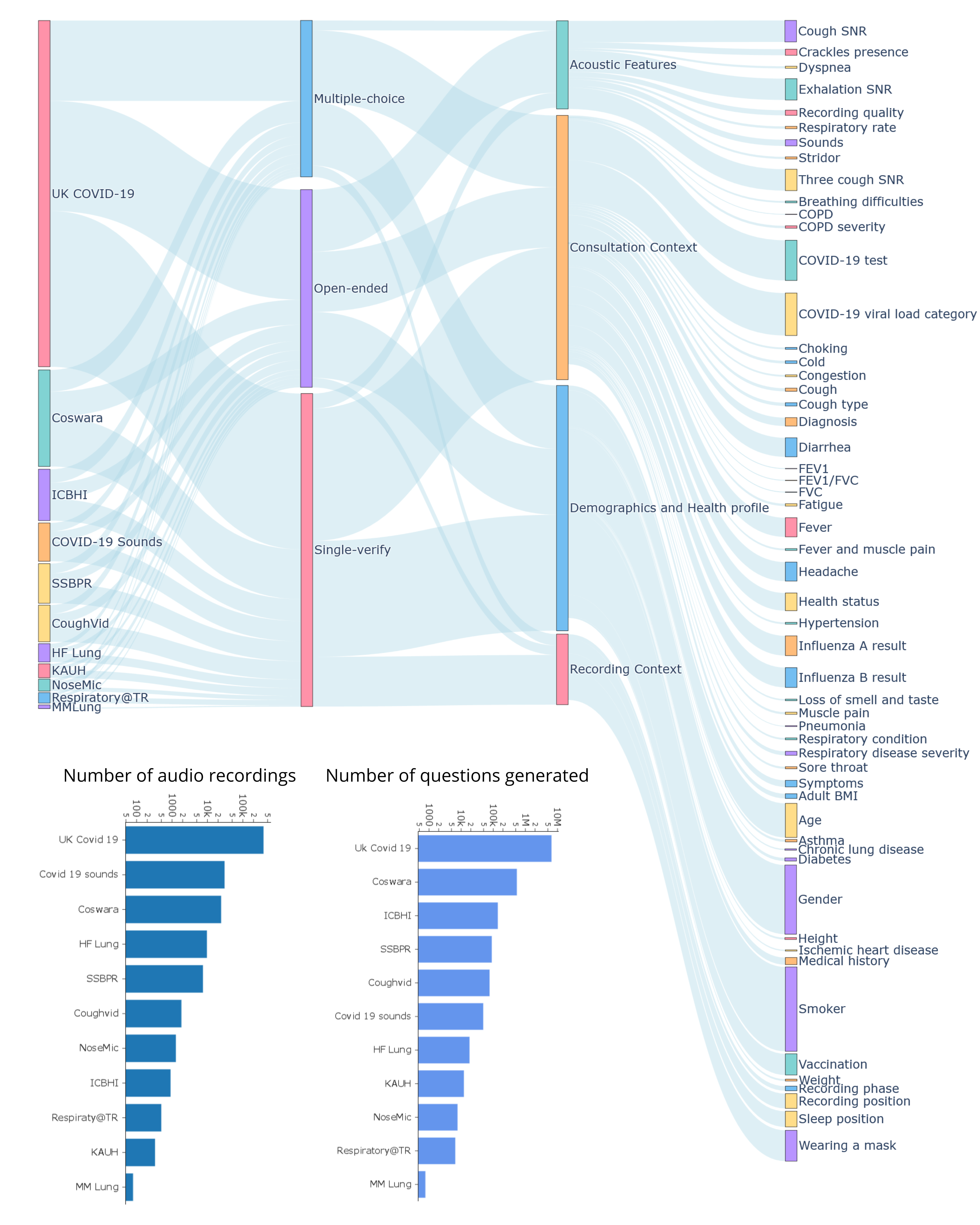}
    \caption{Overview of the RA-QA data flow and the logarithmic distribution of recordings and questions across datasets.}
  \label{fig:sankey}
\end{figure*}

%% file: appendix/tables/mapping_example.tex
% \begin{table*}[h]
% \footnotesize
% \caption{Example of label mapping used to convert dataset-specific annotations into human-readable text. 
% % This mapping ensures consistency across datasets and improves the clarity and interpretability of questions for downstream language models. The example shown is drawn from the \textit{CoughVID} dataset and focuses on the \textit{Diagnosis} attribute.
%   }
%   \label{tab:mapping_example}
%   \centering
%   \begin{tabular}{llll}
%   \toprule
%   \bfseries Dataset & \bfseries Attribute & \bfseries Original label & \bfseries Text mapping\\
%   \midrule
%   \multirow{5}{*}{CoughVID} & \multirow{5}{*}{Diagnosis} & COVID-19 & Positive for COVID-19 \\ 
%    & & healthy\_cough & Healthy cough \\
%    & & lower\_infection & Lower respiratory tract infection\\ 
%    & & obstructive\_disease & Obstructive respiratory disease \\
%    & & upper\_infection & Upper respiratory tract infection \\
%   \bottomrule
%   \end{tabular}
% \end{table*}

\begin{table}[h]
\footnotesize
\caption{Example of label mapping used to convert dataset-specific annotations into human-readable text. 
% This mapping ensures consistency across datasets and improves the clarity and interpretability of questions for downstream language models. The example shown is drawn from the \textit{CoughVID} dataset and focuses on the \textit{Diagnosis} attribute.
  }
  \label{tab:mapping_example}
  \centering
  \begin{tabular}{ll}
  \toprule
  \bfseries Original label & \bfseries Text mapping\\
  \midrule
  COVID-19 & Positive for COVID-19 \\ 
   healthy\_cough & Healthy cough \\
    lower\_infection & Lower respiratory tract infection\\ 
   obstructive\_disease & Obstructive respiratory disease \\
   upper\_infection & Upper respiratory tract infection \\
  \bottomrule
  \end{tabular}
\end{table}

%% file: appendix/tables/example_of_questions.tex
\begin{table*}[p]
  \caption{Examples of RA-QA question templates across open-ended, multiple-choice, and single-verify formats, illustrating consistent attribute-specific question generation.}
  \label{tab:example_of_questions}
  \centering
  \footnotesize
\begin{tabular}{llp{8cm}}
\toprule
\textbf{Question type} & \textbf{Attribute} & \textbf{Example question} \\
\midrule
\multirow{4}{*}{Open-ended}
 & Age & What is the age of the patient? \\
 & Covid test & What is the status of the COVID test for the patient? \\
 & Smoker & What is the patient's smoking history? \\
 & Sleeping position & What is the patient's sleeping position? \\
\midrule
\multirow{5}{*}{Multiple-choice}
 & Gender & What is the gender of the patient? Choose from: ''Female'', ''Male''. \\
 & Recording phase & When was the recording taken? Choose from: ''Before running'', ''After running''. \\
 & Smoking habits & What is the patient's smoking history? Choose from: ''1 to 10 cigarettes per day'', ''11 to 20 cigarettes per day'', ''21+ cigarettes per day'', ''E-cigarette use'', ''Ex-smoker'', ''Less than once a week'', ''Never smoked''. \\
 & Health status & What is the current respiratory health status of the patient? Choose from: ''Healthy'', ''No respiratory illness'', ''Positive but moderate'', ''Fully recovered'', ''Positive but mild'', ''Positive but asymptomatic''. \\
 & Recording quality & What is the quality of the recording? Choose from: ''Good'', ''Ok'', ''Poor''. \\
\midrule
\multirow{5}{*}{Single-verify}
 & Chronic lung disease & Does the patient suffer from chronic lung disease? \\
 & Fatigue & Does the patient have fatigue? \\
 & Recording position & Was the recorder positioned on the posterior right side during the recording? \\
 & Wheezes presence & Are wheezes present in the recording? \\
 & Respiratory disease severity & Is the patient's respiratory disease severe? \\
\bottomrule
\end{tabular}
\end{table*}

%% file: appendix/tables/json_template.tex
\begin{figure*}[p]
\footnotesize
\centering
\lstset{style=jsonstyle} % or style=json if that's what you use
\begin{lstlisting}
[
  {
    "question": "What is the diagnosis of the patient?",
    "question_type": "open-ended"
  },
  {
    "question": "What is the diagnosis for the patient? Choose from: \"Positive for COVID-19\", \"Healthy cough\", \"Lower respiratory tract infection\", \"Obstructive disease\", \"Upper respiratory tract infection\"",
    "question_type": "multiple-choice",
    "covered_labels": [
      "Positive for COVID-19",
      "Healthy cough",
      "Lower respiratory tract infection",
      "Obstructive disease",
      "Upper respiratory tract infection"
    ]
  },
  {
    "question": "Is the diagnosis for the patient positive for COVID-19?",
    "question_type": "single-verify",
    "covered_labels": ["Positive for COVID-19"]
  }
]
\end{lstlisting}
\vspace{-0.25em}
\caption{Example JSON templates used for question generation in RA-QA. The templates illustrate the three supported question types: open-ended, multiple-choice with explicit answer options, and single-verify binary questions.}
\label{lst:json_template_example}
\end{figure*}

%% file: appendix/clinician.tex
\section{Clinician Review Survey}
\label{app:clinician-review}

QA pairs were first formulated from the available metadata fields and then refined through a structured clinical review. The review form presented each proposed QA template together with its candidate answer space, enabling the clinician to assess whether the question formulation was clinically meaningful, whether the terminology was appropriate, and whether the proposed answers captured the intended metadata information. The resulting feedback was incorporated into the final template design to improve clinical clarity, terminology, and consistency.

For each item, the clinician selected one of the following decisions: \emph{Keep as is}, \emph{Minor wording issue}, \emph{Revise possible answers}, \emph{Major rewrite}, or \emph{Remove}. Free-text comments were also collected and used to refine the final wording and answer options. 

%% file: appendix/baselines.tex
\section{Baseline Details}\label{apd:baselines}
% \gb{add more details about models parameter and training}
This appendix provides additional details on the baseline families used in Section~\ref{sec:benchmarking}.
% The baselines are chosen to cover three increasingly informed settings: (i) \emph{label-only} references that ignore inputs, (ii) \emph{classical audio-only} models that use acoustics but not the question, and (iii) a \emph{general multimodal} audio-language model that conditions on the paired $(x,q)$ input. Together, they clarify how much of RA-QA can be solved by label priors, by respiratory acoustics alone, and by out-of-the-box audio-language transfer.\\\\
% \textbf{Label-only baselines (Random / Majority)}\label{apd:baseline_labelonly}
% The \textbf{random} baseline samples uniformly from the valid label set for each target, providing a chance-level reference (particularly informative when labels are balanced). The \textbf{majority} baseline predicts the most frequent training label, computed per attribute (and per question type when applicable), capturing the effect of class imbalance. These baselines ignore both the audio recording and the question text.\\\\
\textbf{Audio-only baselines SVM and OPERA-MLP probes}. To isolate what is predictable from respiratory acoustics without question conditioning, we include two families of audio-only baselines. First, we train a separate \textbf{linear SVM} for each target attribute using a one-vs-rest strategy for multiclass settings~\cite{b29}. All recordings are resampled to 16~kHz mono and converted to 128-bin log-Mel spectrograms with hop length 512 and window length 400. Spectrograms are padded or truncated to a fixed number of frames and flattened into a single feature vector for classification.\\

Second, we train audio-only MLP probes on frozen representations extracted from OPERA encoders~\cite{b6}. OPERA represents one of the most advanced respiratory-specific audio encoding frameworks, providing pretrained encoders tailored to respiratory sound analysis rather than generic audio recognition. We consider OPERA-CT, a contrastive Transformer encoder; OPERA-GT, a generatively pretrained Transformer encoder; and OPERA-CE, a lighter contrastive CNN-based encoder, separately. For each variant, the OPERA encoder is kept frozen and used only as a fixed feature extractor, while a lightweight MLP prediction head is trained on top of the extracted embeddings.
 These baselines estimate the extent to which each OPERA representation supports the considered RA-QA targets from audio alone. Since they do not receive the question text, they are trained separately for each attribute and question type, and therefore should be interpreted as audio-only reference baselines rather than complete RA-QA systems.\\

\textbf{OPERA-based multimodal classifier baselines}\label{apd:baseline_multimodal_classifier}
To evaluate explicit question-audio conditioning, we train late-fusion multimodal classifiers based on frozen OPERA audio representations. For each OPERA variant, namely OPERA-CT, OPERA-GT, and OPERA-CE, the respiratory recording is encoded into an audio embedding, while the corresponding question is encoded with a fixed text encoder. The two embeddings are concatenated and passed to an MLP prediction head.

For discriminative targets, the model uses a classification head, while for regressive targets it uses a scalar regression head. In the global setting, we use a shared fused question-audio representation with task-specific output heads for discriminative and regressive samples. These classifiers allow us to assess how explicit question conditioning changes predictive behaviour compared with audio-only baselines, while keeping the audio encoder fixed.\\

\textbf{General audio-language baselines: Qwen2.5-Omni and Audio Flamingo 3}\label{apd:baseline_general_audio_language}
We evaluate Qwen2.5-Omni~\cite{b33} and Audio Flamingo 3~\cite{b34} as representative general-purpose audio-language models to test zero-shot transfer to clinically grounded respiratory QA. These models provide strong recent examples of broadly trained audio-language systems: Qwen2.5-Omni follows an omni-modal design supporting audio together with other modalities, whereas Audio Flamingo 3 is designed for general audio understanding and reasoning across speech, sound, and music. Their inclusion therefore tests whether general audio-language alignment is sufficient to recover respiratory attributes without domain-specific adaptation. Unlike the attribute-specific audio-only and classifier baselines, these models are evaluated on the paired input $(x,q)$ and generate free-form textual answers. We use them without respiratory-specific fine-tuning and prompt each model with the respiratory audio and the associated question, using the same answer-parsing and evaluation protocol adopted for the other generative baselines.

% The generated answers are then normalized and parsed for task-level scoring, while BERTScore is used to measure semantic overlap with the reference answers.

% \input{figures/figures_tex/appendix_supplementary/pengi}

%% file: appendix/exp_setting.tex
\section{Experimental setting} \label{apd:exp_setting}
\input{appendix/tables/exp_setting}
Table~\ref{tab:exp_setting} reports the statistics of the respiratory-audio RA-QA subset used in the experimental setting. This subset was selected to retain the heterogeneity of the broader collection while defining a tractable and consistent benchmark for model training and evaluation. It includes 1,402 patients, 18,750 unique audio recordings, and 148,037 examples. 
% The data are originally organized at the patient level, where each JSON object may contain multiple recordings associated with the same question and answer. For the experimental setup, these patient-level objects are expanded over their corresponding audio files, so that each individual question-audio-answer triplet is treated as an independent example.

% The subset is intentionally heterogeneous. It combines datasets collected under different acquisition conditions and with different recording devices, audio types, patient populations, annotation schemas, and available metadata. For example, the number of patients and recordings varies substantially across sources, from smaller clinically focused datasets such as HF Lung and Respiratory@TR to larger collections such as ICBHI and SSBPR. 
% Similarly, the number of questions, attributes, and unique answers differs across datasets, reflecting variation in the type and granularity of information available in each source. 
This design allows the experiments to evaluate whether respiratory-audio QA models can generalize across diverse data sources, question formulations, answer spaces, and audio acquisition scenarios.

%% file: appendix/tables/exp_setting.tex
\begin{table*}[htbp]
\caption{Overview statistics of the respiratory-audio RA-QA subset used in the experimental setting.}\vspace{-5pt}
\label{tab:exp_setting}
\centering
\setlength{\tabcolsep}{3pt}

\resizebox{\textwidth}{!}{%
\begin{tabular}{lrrrrrrrrrrrr}
\toprule
 & \textbf{KAUH} & \textbf{Coswara} & \textbf{CoughVID} & \textbf{COVID-19 Sounds} & \textbf{HF Lung} & \textbf{ICBHI} & \textbf{MM Lung} & \textbf{NoseMic} & \textbf{SSBPR} & \textbf{UK COVID-19} & \textbf{Resp@TR} & \textbf{All} \\
\midrule
Number of Patients & 76 & 222 & 348 & 144 & 75 & 111 & 33 & 14 & 19 & 344 & 16 & 1402 \\
Number of Unique Audio Samples & 228 & 1998 & 348 & 435 & 75 & 6676 & 66 & 1129 & 7258 & 345 & 192 & 18750 \\
\midrule
Mean Question Length (words) & 11.05 & 8.58 & 8.38 & 8.68 & 10.09 & 10.16 & 8.63 & 7.90 & 9.23 & 10.88 & 10.71 & 9.59 \\
Number of Unique Questions & 30 & 17 & 30 & 47 & 9 & 25 & 8 & 6 & 12 & 30 & 7 & 168 \\
Number of Attributes & 5 & 4 & 7 & 6 & 2 & 9 & 5 & 2 & 2 & 11 & 1 & 54 \\
\midrule
Mean Answer Length (words) & 6.83 & 8.42 & 6.09 & 8.66 & 8.00 & 5.85 & 7.19 & 7.27 & 7.08 & 5.62 & 9.07 & 6.64 \\
Number of Unique Answers & 133 & 121 & 152 & 325 & 20 & 222 & 125 & 570 & 32 & 1423 & 17 & 2915 \\
\midrule
\textbf{QA Pairs} & 1887 & 9378 & 1901 & 5352 & 345 & 69056 & 474 & 6298 & 50040 & 1962 & 1344 & 148037 \\
\bottomrule
\end{tabular}%
}

\vspace{2pt}
\begin{flushleft}
\footnotesize
Each QA pair corresponds to an individual question-audio-answer triplet obtained by expanding patient-level objects over their associated audio recordings.
\end{flushleft}

\end{table*}

%% file: appendix/results.tex
\section{Comprehensive Benchmark Results} \label{sec:benchmark_results}
We report both task-level scores (Accuracy for discriminative targets; MAE for regressive targets) and text-level semantic fidelity (BERTScore). This decomposition is intentional: under RA-QA's heterogeneity, high semantic similarity can coexist with low task-level correctness, particularly for open-ended questions where plausible text may not match the required canonical label or numeric target. The per-dataset breakdown further illustrates substantial variation across recording conditions and dataset-specific label definitions, reinforcing the need to evaluate robustness across sources and the alignment between natural-language generations and clinically relevant targets.\\

\noindent \textbf{CaReAQA-style model}. Tables~\ref{tab:careaqa_accuracy}-\ref{tab:careaqa_bert} summarize the CaReAQA-style generative baseline. Across datasets, CaReAQA typically attains high BERTScore, yet its task-level performance varies widely by dataset and question format, exemplifying the semantic-task-level mismatch that RA-QA is designed to expose in heterogeneous settings.\\

\noindent \textbf{Multimodal classifier}. Tables~\ref{tab:multimodal_accuracy}-\ref{tab:multimodal_bert} report results for the multimodal classifier baseline. Two recurring properties emerge: (i) performance is strongly format-dependent, with single-verify often easier than open-ended or multiple-choice due to the constrained decision interface; and (ii) high BERTScore can coincide with modest Accuracy/MAE, indicating that semantically plausible outputs do not necessarily match the canonical label or numeric value. Overall, even this question-conditioned baseline exhibits substantial variation across datasets and acquisition conditions, highlighting the challenge of generalization under heterogeneous respiratory audio sources.\\

\noindent \textbf{General audio-language models}. 
Tables~\ref{tab:qwen_bert}-\ref{tab:qwen_accuracy} -\ref{tab:flamingo_bert}-\ref{tab:flamingo_accuracy} report the zero-shot results for Qwen2.5-Omni and Audio Flamingo 3. Both models show limited transfer to clinically grounded respiratory QA, with weak task-level performance especially on open-ended discriminative questions. Their relatively higher scores on multiple-choice questions suggest that constrained answer formats are easier for instruction-following audio-language models than free-form attribute prediction. However, the models still frequently fail to recover the correct canonical label or numeric target. As with the other generative baselines, BERTScore can remain moderate or high even when task-level correctness is low, reflecting fluent but generic or non-specific responses. These results indicate that generic audio-language transfer is insufficient for RA-QA and motivate models trained to align respiratory audio evidence, question semantics, and task-level prediction.
\input{appendix/tables/careaqa_accuracy}  
\input{appendix/tables/careaqa_bert}  
\input{appendix/tables/multimodal_accuracy}

\input{appendix/tables/multimodal_bert}  
\input{appendix/tables/qwen_accuracy}

\input{appendix/tables/qwen_bert}

\input{appendix/tables/flamingo_accuracy}  
\input{appendix/tables/flamingo_bert}

%% file: appendix/tables/careaqa_accuracy.tex
\begin{table*}[p]
  \caption{CaReAQA-style baseline: task-level performance per dataset. We report Accuracy for discriminative tasks (Single-verify / Open-ended / Multiple-choice) and MAE for regressive tasks (Open-ended only). All values are rounded to two decimals.}
  \label{tab:careaqa_accuracy}
  \centering
  \begin{tabular}{lcccc}
    \toprule
    Dataset & \multicolumn{3}{c}{\textbf{Discriminative tasks (Accuracy)}} & \textbf{Regressive tasks (MAE)} \\ \midrule
           & \textbf{Single-verify} & \textbf{Open-ended} & \textbf{Multiple-choice} & \textbf{Open-ended }\\
    \midrule
    Coswara          & 0.60 & 0.89 & 1.00 & -- \\
    CoughVID         & 0.47 & 0.30 & 0.40 & -- \\
    COVID-19 Sounds  & 0.58 & 0.38 & 0.50 & -- \\
    HF Lung          & 0.73 & 0.53 & 0.53 & -- \\
    ICBHI            & 0.24 & 0.07 & 0.52 & -- \\
    KAUH             & 0.49 & 0.06 & 0.35 & -- \\
    MM Lung          & 0.50 & 0.00 & 0.10 & 0.83 \\
    NoseMic          & 0.50 & 0.47 & 0.54 & 2.03 \\
    SSBPR            & 0.37 & 0.00 & 0.66 & -- \\
    Respiratory@TR    & 0.36 & 0.42 & 0.80 & -- \\
        UK COVID-19      & 0.37 & 0.32 & 0.43 & 0.85 \\

    \bottomrule
  \end{tabular}
\end{table*}

%% file: appendix/tables/careaqa_bert.tex
\begin{table*}[p]
  \caption{CaReAQA-style baseline: text-level semantic fidelity per dataset. We report BERTScore for discriminative tasks (Single-verify / Open-ended / Multiple-choice) and for regressive tasks (Open-ended only). All values are rounded to two decimals.}
  \label{tab:careaqa_bert}
  \centering
  \begin{tabular}{lcccc}
    \toprule
    Dataset & \multicolumn{3}{c}{\textbf{Discriminative tasks (BERTScore)}} & \textbf{Regressive tasks (BERTScore)} \\ \midrule
           & \textbf{Single-verify} & \textbf{Open-ended} & \textbf{Multiple-choice} & \textbf{Open-ended }\\
    \midrule
    Coswara          & 0.97 & 1.00 & 1.00 & -- \\%0.94 \\
    CoughVID         & 0.97 & 0.96 & 0.96 & -- \\%0.94 \\
    COVID-19 Sounds  & 0.93 & 0.94 & 0.94 & -- \\%0.97 \\
    HF Lung          & 0.99 & 1.00 & 0.94 & -- \\%0.00 \\
    ICBHI            & 0.98 & 0.93 & 0.96 & -- \\%0.97 \\
    KAUH             & 0.98 & 0.96 & 0.97 & -- \\%0.94 \\
    MM Lung          & 0.99 & 0.94 & 1.00 & 0.99 \\
    NoseMic          & 0.98 & 0.99 & 1.00 & 0.99 \\
    Respiratory@TR   & 0.98 & 0.99 & 0.98 & -- \\%0.00 \\
    SSBPR            & 1.00 & 0.98 & 0.97 & -- \\%0.00 \\
    UK COVID-19      & 0.99 & 0.98 & 0.99 & 0.96 \\
    \bottomrule
  \end{tabular}
\end{table*}

%% file: appendix/tables/multimodal_accuracy.tex
\begin{table*}[p]
  \caption{Global multimodal classifier baseline: task-level performance per dataset.  We report Accuracy for discriminative tasks (Single-verify / Open-ended / Multiple-choice) and MAE for regressive tasks (Open-ended only). All values are rounded to two decimals.}
  \label{tab:multimodal_accuracy}
  \centering
  \begin{tabular}{lcccc}
    \toprule
    Dataset & \multicolumn{3}{c}{\textbf{Discriminative tasks (Accuracy)}} & \textbf{Regressive tasks (MAE)} \\ \midrule
           & \textbf{Single-verify} & \textbf{Open-ended} & \textbf{Multiple-choice} & \textbf{Open-ended }\\
    \midrule
    Coswara          & 0.01 & 0.04 & 0.03 & -- \\
    CoughVID         & 0.04 & 0.17 & 0.25 & -- \\
    COVID-19 Sounds  & 0.01 & 0.02 & 0.01 & -- \\
    HF Lung          & 0.13 & 0.17 & 0.04 & -- \\
    ICBHI            & 0.19 & 0.01 & 0.01 & -- \\
    KAUH             & 0.10 & 0.02 & 0.04 & -- \\
    MM Lung          & 0.73 & 1.00 & 0.89 & 0.53 \\
    NoseMic          & 0.01 & 0.01 & 0.01 & 2.38 \\
    SSBPR            & 0.45 & 0.19 & 0.21 & -- \\
        Respiratory@TR    & 0.29 & 0.02 & 0.02 & -- \\
    UK COVID-19      & 0.01 & 0.02 & 0.03 & 0.06 \\
    \bottomrule
  \end{tabular}
\end{table*}

%% file: appendix/tables/multimodal_bert.tex
\begin{table*}[p]
  \caption{Global multimodal classifier baseline: text-level semantic fidelity per dataset. We report BERTScore for discriminative tasks (Single-verify / Open-ended / Multiple-choice) and for regressive tasks (Open-ended only). All values are rounded to two decimals.}
  \label{tab:multimodal_bert}
  \centering
  \begin{tabular}{lcccc}
    \toprule
    Dataset & \multicolumn{3}{c}{\textbf{Discriminative tasks (BERTScore)}} & \textbf{Regressive tasks (BERTScore)} \\\midrule
           & \textbf{Single-verify} & \textbf{Open-ended} & \textbf{Multiple-choice} & \textbf{Open-ended }\\
    \midrule
    Coswara          & 0.84 & 0.81 & 0.85 & -- \\
    CoughVID         & 0.85 & 0.84 & 0.89 & -- \\
    COVID-19 Sounds  & 0.84 & 0.83 & 0.84 & -- \\
    HF Lung          & 0.82 & 0.88 & 0.86 & -- \\
    ICBHI            & 0.84 & 0.84 & 0.93 & -- \\
    KAUH             & 0.83 & 0.83 & 0.81 & -- \\
    MM Lung          & 0.84 & 0.83 & 0.80 & 0.86 \\
    NoseMic          & 0.83 & 0.85 & 0.88 & 0.85 \\
    SSBPR            & 0.84 & 0.85 & 0.89 & -- \\
        Respiratory@TR    & 0.84 & 0.83 & 0.96 & -- \\

    UK COVID-19      & 0.87 & 0.83 & 0.88 & 0.84 \\
    \bottomrule
  \end{tabular}
\end{table*}

%% file: appendix/tables/qwen_accuracy.tex
% \begin{table*}[p]
%   \caption{Pengi zero-shot baseline: task-level performance per dataset. We report Accuracy for discriminative tasks (Single-verify / Open-ended / Multiple-choice). Regression is marked as NA when Pengi does not yield usable numeric outputs under our parsing protocol. All values are rounded to two decimals.}
%   \label{tab:pengi_macro}
%   \centering
%   \begin{tabular}{lcccc}
%     \toprule
%     Dataset & \multicolumn{3}{c}{\textbf{Discriminative tasks (Accuracy)}} & \textbf{Regressive tasks (MAE)} \\ \midrule
%            & \textbf{Single-verify} & \textbf{Open-ended} & \textbf{Multiple-choice} & \textbf{Open-ended }\\
%     \midrule
%     Coswara         & 0.00 & 0.00 & 0.00 & NA \\
%     CoughVID        & 0.00 & 0.01 & 0.00 & NA \\
%     COVID-19 Sounds & 0.00 & 0.00 & 0.00 & NA \\
%     HF Lung         & 0.44 & 0.00 & 0.00 & NA \\
%     ICBHI           & 0.00 & 0.00 & 0.00 & NA \\
%     KAUH            & 0.01 & 0.00 & 0.00 & NA \\
%     MM Lung         & 0.02 & 0.03 & 0.00 & NA \\
%     NoseMic         & 0.00 & 0.00 & 0.00 & NA \\
%     Respiratory@TR  & 0.02 & 0.00 & 0.00 & NA \\
%     SSBPR           & 0.02 & 0.01 & 0.00 & NA \\
%     UK COVID-19     & 0.06 & 0.00 & 0.00 & NA \\
%     \bottomrule
%   \end{tabular}
% \end{table*}

\begin{table*}[p]
\caption{Qwen2.5-Omni zero-shot baseline: task-level performance per dataset. We report Accuracy for discriminative tasks (Single-verify / Open-ended / Multiple-choice) and MAE for regressive tasks (Open-ended only). All values are rounded to two decimals.}
\label{tab:qwen_accuracy}
\centering
\begin{tabular}{lcccc}
\toprule
Dataset & \multicolumn{3}{c}{\textbf{Discriminative tasks (Accuracy)}} & \textbf{Regressive tasks (MAE)} \\
\midrule
& \textbf{Single-verify} & \textbf{Open-ended} & \textbf{Multiple-choice} & \textbf{Open-ended} \\
\midrule
 Coswara          & 0.39 & 0.39 & 0.28 & -- \\
    CoughVID         & 0.39 & 0.13 & 0.35 & -- \\
    COVID-19 Sounds  & 0.76 & 0.28 & 0.49 & -- \\
    HF Lung          & 0.59 & 0.00 & 0.53 & -- \\
    ICBHI            & 0.20 & 0.12 & 0.22 & -- \\
    KAUH             & 0.22 & 0.15 & 0.24 & -- \\
    MM Lung          & 0.71 & 0.50 & 1.00 & 0.68 \\
    NoseMic          & 0.25 & 0.00 & 1.00 & 6.14 \\
    Respiratory@TR   & 0.40 & 0.00 & 0.20 & -- \\
    SSBPR            & 0.12 & 0.00 & 1.00 & -- \\
    UK COVID-19      & 0.59 & 0.04 & 0.21 & 1.33 \\
\bottomrule
\end{tabular}
\end{table*}

%% file: appendix/tables/qwen_bert.tex
\begin{table*}[p]
  \caption{Qwen2.5-Omni zero-shot baseline: text-level semantic fidelity per dataset. We report BERTScore for discriminative tasks (Single-verify / Open-ended / Multiple-choice) and for regressive tasks (Open-ended only, when applicable). All values are rounded to two decimals.}
\label{tab:qwen_bert}
\centering
\begin{tabular}{lcccc}
\toprule
Dataset & \multicolumn{3}{c}{\textbf{Discriminative tasks (BERTScore)}} & \textbf{Regressive tasks (BERTScore)} \\
\midrule
& \textbf{Single-verify} & \textbf{Open-ended} & \textbf{Multiple-choice} & \textbf{Open-ended} \\
\midrule
Coswara & 0.90 & 0.93 & 0.96 & -- \\
CoughVID & 0.90 & 0.92 & 0.92 & -- \\
COVID-19 Sounds & 0.90 & 0.91 & 0.93 & -- \\
HF Lung & 0.92 & 0.88 & 0.91 & -- \\
ICBHI & 0.86 & 0.89 & 0.96 & -- \\
KAUH & 0.86 & 0.90 & 0.94 & -- \\
MM Lung & 0.87 & 0.89 & 1.00 & 0.94 \\
NoseMic & 0.87 & 0.93 & 1.00 & 0.96 \\
Respiratory@TR & 0.90 & 0.92 & 0.96 & -- \\%0.86 \\
SSBPR & 0.88 & 0.93 & 1.00 & -- \\
UK COVID-19 & 0.94 & 0.90 & 0.97 & 0.85 \\
\bottomrule
\end{tabular}
\end{table*}

%% file: appendix/tables/flamingo_accuracy.tex
\begin{table*}[p]
\caption{Audio Flamingo 3 zero-shot baseline: task-level performance per dataset. We report Accuracy for discriminative tasks (Single-verify / Open-ended / Multiple-choice) and MAE for regressive tasks (Open-ended only). All values are rounded to two decimals.}\label{tab:flamingo_accuracy}
\centering
\begin{tabular}{lcccc}
\toprule
Dataset & \multicolumn{3}{c}{\textbf{Discriminative tasks (Accuracy)}} & \textbf{Regressive tasks (MAE)} \\
\midrule
& \textbf{Single-verify} & \textbf{Open-ended} & \textbf{Multiple-choice} & \textbf{Open-ended} \\
\midrule
    Coswara          & 0.42 & 0.03 & 0.22 & -- \\
    CoughVID         & 0.53 & 0.03 & 0.39 & -- \\
    COVID-19 Sounds  & 0.43 & 0.12 & 0.36 & -- \\
    HF Lung          & 0.50 & 0.00 & 0.43 & -- \\
    ICBHI            & 0.21 & 0.01 & 0.21 & -- \\
    KAUH             & 0.24 & 0.05 & 0.24 & -- \\
    MM Lung          & 0.83 & 0.00 & 0.50 & 0.07 \\
    NoseMic          & 0.50 & 0.00 & 1.00 & 1.86 \\
    Respiratory@TR   & 0.31 & 0.00 & 0.60 & -- \\
    SSBPR            & 0.12 & 0.00 & 1.00 & -- \\
    UK COVID-19      & 0.32 & 0.01 & 0.20 & 2.30 \\
\bottomrule
\end{tabular}
\end{table*}

%% file: appendix/tables/flamingo_bert.tex
\begin{table*}[p]
  \caption{Audio Flamingo 3 zero-shot baseline: text-level semantic fidelity per dataset. We report BERTScore for discriminative tasks (Single-verify / Open-ended / Multiple-choice) and for regressive tasks (Open-ended only, when applicable). All values are rounded to two decimals.}
\label{tab:flamingo_bert}
\centering
\begin{tabular}{lcccc}
\toprule
Dataset & \multicolumn{3}{c}{\textbf{Discriminative tasks (BERTScore)}} & \textbf{Regressive tasks (BERTScore)} \\
\midrule
& \textbf{Single-verify} & \textbf{Open-ended} & \textbf{Multiple-choice} & \textbf{Open-ended} \\
\midrule
Coswara & 0.91 & 0.90 & 0.96 & -- \\
CoughVID & 0.90 & 0.91 & 0.92 & -- \\
COVID-19 Sounds & 0.92 & 0.92 & 0.92 & -- \\
HF Lung & 0.97 & 0.86 & 0.92 & -- \\
ICBHI & 0.93 & 0.89 & 0.94 & -- \\
KAUH & 0.87 & 0.87 & 0.94 & -- \\
MM Lung & 0.95 & 0.87 & 1.00 & 0.90 \\
NoseMic & 0.96 & 0.91 & 1.00 & 0.93 \\
Respiratory@TR & 0.88 & 0.89 & 0.98 & -- \\
SSBPR & 0.93 & 0.90 & 1.00 & -- \\
UK COVID-19 & 0.94 & 0.89 & 0.96 & 0.83 \\
\bottomrule
\end{tabular}
\end{table*}